 \documentclass[final,3p,times]{elsarticle}

\usepackage{amssymb}
\usepackage{amsmath}
\usepackage{amsthm}

\usepackage{lipsum}
\usepackage{graphicx}
\usepackage{dcolumn}
\usepackage{bm}
\usepackage{multirow}
\usepackage{comment}
\usepackage{subfigure}
\usepackage{soul}

\usepackage[usenames,dvipsnames]{color}
\usepackage{hyperref}

\usepackage[mathlines]{lineno}

\usepackage{lineno}
\usepackage{setspace}

\usepackage{verbatim} 

\DeclareGraphicsExtensions{.pdf,.png,.jpg}

\biboptions{sort&compress}

\journal{Nuclear~Instruments~and~Methods~A}

\begin{document}

\newcommand\JOcom[1]{{\textcolor{red}{#1}}}

\newcommand{\bmd}{$\beta$}
\newcommand{\bpd}{$\beta^{+}$}
\newcommand{\bb}{$\beta\beta$}
\newcommand{\nbb}{\ensuremath{\nu\beta\beta}}
\newcommand{\Te}{$^{130}$Te}
\newcommand{\Se}{$^{82}$Se}
\newcommand{\B}{$^{8}$B}
\newcommand{\Bten}{$^{10}$B}
\newcommand{\Cten}{$^{10}$C}
\newcommand{\C}{$^{10}$C}



\large

\title{Separating Double-Beta Decay Events from Solar Neutrino
Interactions in a Kiloton-Scale Liquid Scintillator Detector By Fast Timing}
\author[UChicago]{Andrey Elagin\corref{cor1}}
\cortext[cor1]{Corresponding Author: \href{mailto:elagin@hep.uchicago.edu}{\tt{elagin@hep.uchicago.edu}}}
\author[UChicago]{Henry J. Frisch}
\author[UCLA]{Brian Naranjo}
\author[MIT]{Jonathan Ouellet}
\author[MIT]{Lindley Winslow}
\author[MIT]{Taritree Wongjirad}

\address[UChicago]{ Enrico Fermi Institute, University of Chicago, Chicago, IL, 60637 }
\address[UCLA]{ University of California, Los Angeles, CA, 90024 }
\address[MIT]{ Massachusetts Institute of Technology, Cambridge, MA 02139 }

\begin{abstract}
We present a technique for separating nuclear double beta decay
(\bb-decay) events from background neutrino interactions due to
\B~decays in the sun.  This background becomes dominant in a
kiloton-scale liquid-scintillator detector deep underground and is
usually considered as irreducible due to an overlap in deposited
energy with the signal.  However, electrons from 0\nbb-decay often
exceed the Cherenkov threshold in liquid scintillator, producing
photons that are prompt and correlated in direction with the
initial electron direction. The use of large-area fast photodetectors
allows some separation of these prompt photons from delayed isotropic
scintillation light and, thus, the possibility of reconstructing the
event topology.  Using a simulation of a 6.5~m radius liquid
scintillator detector with 100~ps resolution photodetectors, we show
that a spherical harmonics analysis of early-arrival light can
discriminate between 0\nbb-decay signal and
\B~ solar neutrino background events on a statistical basis. 
Good separation will require the development of a slow scintillator 
with a 5 nsec risetime.


\end{abstract}

\date{\today}

\maketitle

\clearpage
\tableofcontents

\clearpage

\section{Introduction}

The electron, muon, and tau neutrinos are unique among the standard
model fermions in being electrically neutral and orders-of-magnitude
less massive than their standard model charged
partners~\cite{PDG_mass}.  These two properties motivate the
possibility that these neutrinos are `Majorana' rather than `Dirac'
particles, i.e. different from their respective charged partner
leptons by being their own
anti-particle~\cite{Majorana1937,PDG_mass}. In 1939 W. Furry
pointed out that a Majorana nature of the electron neutrino would
allow neutrinoless double-beta decay, in which a nucleus undergoes a
second order $\beta$-decay without producing any neutrinos,
$(Z,A)\rightarrow(Z+2,A)+2\beta^-$~\cite{Furry1939}.  This is in
contrast to the Goeppert-Mayer two-neutrino double beta (2{\nbb})
decay, the second order standard model (SM) \bmd-decay channel in which
lepton number is conserved by the production of two anti-neutrinos,
\mbox{$(Z,A)\rightarrow(Z,A+2)+2\beta+2\bar\nu_e$}~\cite{GoeppertMayer1935}.

The standard mechanism of 0\nbb-decay is parametrized by the
effective Majorana mass, defined as
\mbox{$m_{\beta\beta}\equiv\left|\sum_i U^2_{ei}m_i\right|$}, where
$U_{ei}$ are the elements of the PMNS matrix and $m_i$ are the
neutrino masses~\cite{PDG_mass}. Current half-life limit
translate to a limit on \mbox{$m_{\beta\beta}\lesssim
61-165\,\mathrm{meV}$}~\cite{KamLANDZen2016}.  The next generation of 0\nbb-decay
experiments~\cite{NSACreport} seek to be sensitive enough to
detect or rule out 0\nbb-decay down to \mbox{$m_{\beta\beta}\lesssim
10$~meV}. This will require a detector to instrument roughly a ton of
active isotope with good energy resolution and a near zero background.

Liquid scintillator-based detectors have proven to be a competitive
technology~\cite{KamLANDZen2013} and offer the advantage of
scalability to larger instrumented masses by dissolving larger amounts
of the isotope of interest into the liquid scintillator (LS).  This
may allow scaling to 1~ton or more of isotope using detectors already
in operation \cite{Biller2013}.  In a large LS detector, most
backgrounds can be strongly suppressed through a combination of
filtration of the LS to remove internal contaminants, self-shielding
to minimize the effects of external contaminants, and vetoes to reduce
muon spallation backgrounds. The dominant backgrounds are
the standard model 2\nbb-decay and electron scattering of 
neutrinos from $^8$B decays in the sun.

 In a previous work~\cite{Aberle2014} we have shown that large-area
photo-detectors with timing resolution of $\sim$100~ps can be used to
resolve prompt Cherenkov photons from the slower scintillation signal
in a large LS detector and that the resulting distributions can be fit for
the directions and origin of $\sim$MeV electrons. Here we present a
study of applying this technique to the topological separation of
0\nbb-decay signal and \B~ background using a spherical harmonic
decomposition to analyze the distribution of early (and hence weighted
toward Cherenkov photons) photoelectrons (PEs) as a topological
discriminant.

The organization of the paper is as follows. 
Section~\ref{sec:detector_description} describes
the detector model. Details on event
kinematics and PE timing for signal and background are given in
Section~\ref{sec:kinematics_and_timing}. In
Section~\ref{sec:topology_and_harmonics}, we introduce the spherical
harmonic decomposition and discuss the performance of this analysis
in Section~\ref{sec:performance}. The conclusions are summarized in Section~\ref{sec:conclusions}.

\section{Detector Model}
\label{sec:detector_description}

We use the Geant4-based simulation of Ref.~\cite{Aberle2014}
to model a sphere of 6.5~m radius filled with liquid scintillator. We
consequently limit the discussion of the simulation to a summary of
the most relevant parameters.

The scintillator composition has been chosen to match a KamLAND-like
scintillator\cite{kamland2003}. The composition is 80\% n-dodecane,
20\% pseudocumene and 1.52~g/l PPO with a density of $\rho$ =
0.78~g/ml).  We use the Geant4~default liquid scintillator optical
model, in which optical photons are assigned the group velocity in the
wavelength region of normal dispersion. The attenuation
length\cite{tajimaMaster}, scintillation emission
spectrum\cite{tajimaMaster}, and refractive index\cite{OlegThesis}
include wavelength-dependence. The scintillator light yield is assumed
to be 9030 photons/MeV) with Birks quenching ($kB$ $\approx$
0.1~mm/MeV)\cite{ChrisThesis}. However, we deviate from the
baseline KamLAND case in that the re-emission of absorbed photons in
the scintillator bulk volume and optical scattering, specifically
Rayleigh scattering, have not yet been included. A test simulation
shows that the effect of optical scattering is
negligible~\cite{Aberle2014}.

The technique of using Cherenkov light for topological \B~ background rejection
depends on the inherent time constants that (on average) slow
scintillation light relative to the Cherenkov light for wavelengths
longer than the scintillator absorption cutoff (between
360-370~nm~\cite{ChristophThesis}). The first step in the scintillation
process is the transfer of energy deposited by the primary
particles from the scintillator's solvent to the solute. The time
constant of this energy transfer accounts for a rise time in
scintillation light emission. Because past neutrino experiments were
not highly sensitive to the effect of the scintillation rise time,
there is a lack of accurate measurements of this property. We assume a
rise time of 1.0~ns from a re-analysis of the data in
Ref.~\cite{ChristophThesis} but more detailed studies are needed. 

The decay time constants are determined by the vibrational energy
levels of the solute and are measured to be $\tau_{d1}$ = 6.9~ns and
$\tau_{d2}$ = 8.8~ns with relative weights of 0.87 and 0.13 for the
KamLAND scintillator~\cite{tajimaThesis}. In a detector of this size,
chromatic dispersion, wherein red light traveling faster than blue due to the
wavelength-dependent index of refraction, enhances the separation.

The inner sphere surface is used as the photodetector. It is treated
as fully absorbing with no reflections and with 100\% photocathode
coverage. As in the case of optical scattering,
reflections at the sphere are a small effect that would create a small
tail at longer times and, hence, does not affect the identification of the
early Cherenkov light. The assumed quantum efficiency (QE) is that
of a typical bialkali photocathode (Hamamatsu R7081
PMT~\cite{Hamamatsu_R7081}, see also Ref.~\cite{dctwo}), which is 12\% for 
Cherenkov light and 23\% for scintillation light.
We note that the KamLAND 17-inch PMTs use the same photocathode type with similar
quantum efficiency; photocathodes with higher efficiencies are now
starting to become better understood theoretically and may become
commercially available~\cite{Photonis, Smedley, Cultrera}.  In order to
neglect the effect of the transit-time-spread (TTS) of the
photodetectors, we use a TTS of 100~ps ($\sigma$), which, for example, can be
achieved with large area picosecond photodetectors
(LAPPDs)~\cite{Timing_paper}.
We neglect the (small) threshold effects in the photodetector readout
electronics, spatial resolution of the photoelectron hit positions,
and contributions to time resolution other than the photodetector TTS.

\section{Kinematics and Timing of Signal and Background events}
\label{sec:kinematics_and_timing}

\subsection{Kinematics of the 0\nbb-decay signal}

\begin{figure*}[ht]
  \centering
  \includegraphics[width=0.65\textwidth]{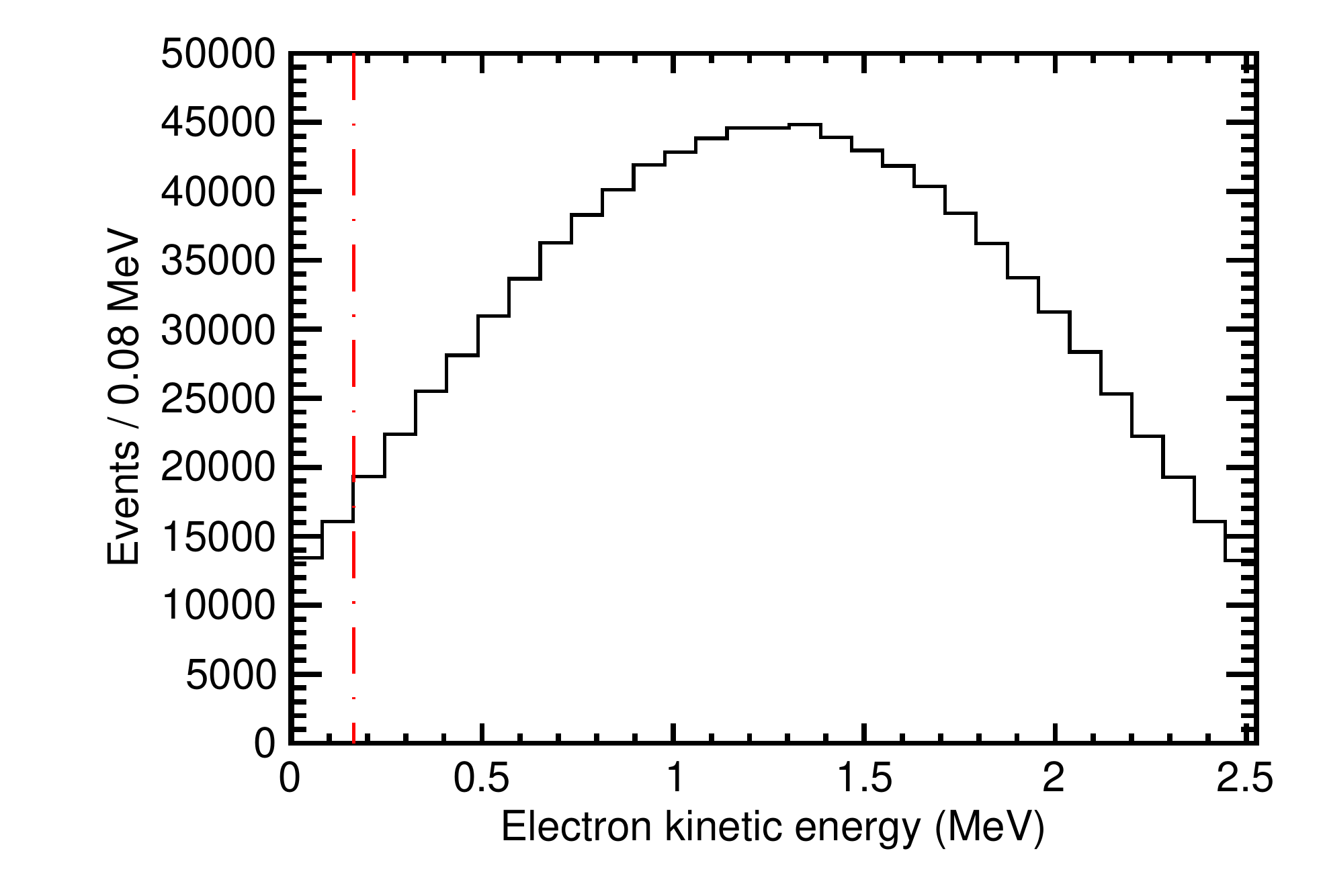}
  \caption{The spectrum in kinetic energy of one of the electrons in 0\nbb-
  decays of \Te~ (endpoint 2.53 MeV). The vertical dashed line
  indicates the Cherenkov threshold in the liquid scintillator of the
  detector model. Single electrons from \B~ solar neutrinos that are
  potential background to the 0\nbb-decay search are close in energy to the
  endpoint and will be above the Cherenkov threshold.}
 \label{fig:Energy_spectrum}
\end{figure*}

We simulate the kinematics of 0\nbb-decay events using a custom Monte
Carlo with momentum and angle-dependent phase space
factors for 0\nbb-decay~\cite{Jenni}.  The spectrum in kinetic energy
of one electron in 0\nbb-decays of \Te~ is shown in
Figure~\ref{fig:Energy_spectrum}.

\begin{figure*}[ht]
  \centering
  \includegraphics[width=0.49\textwidth]{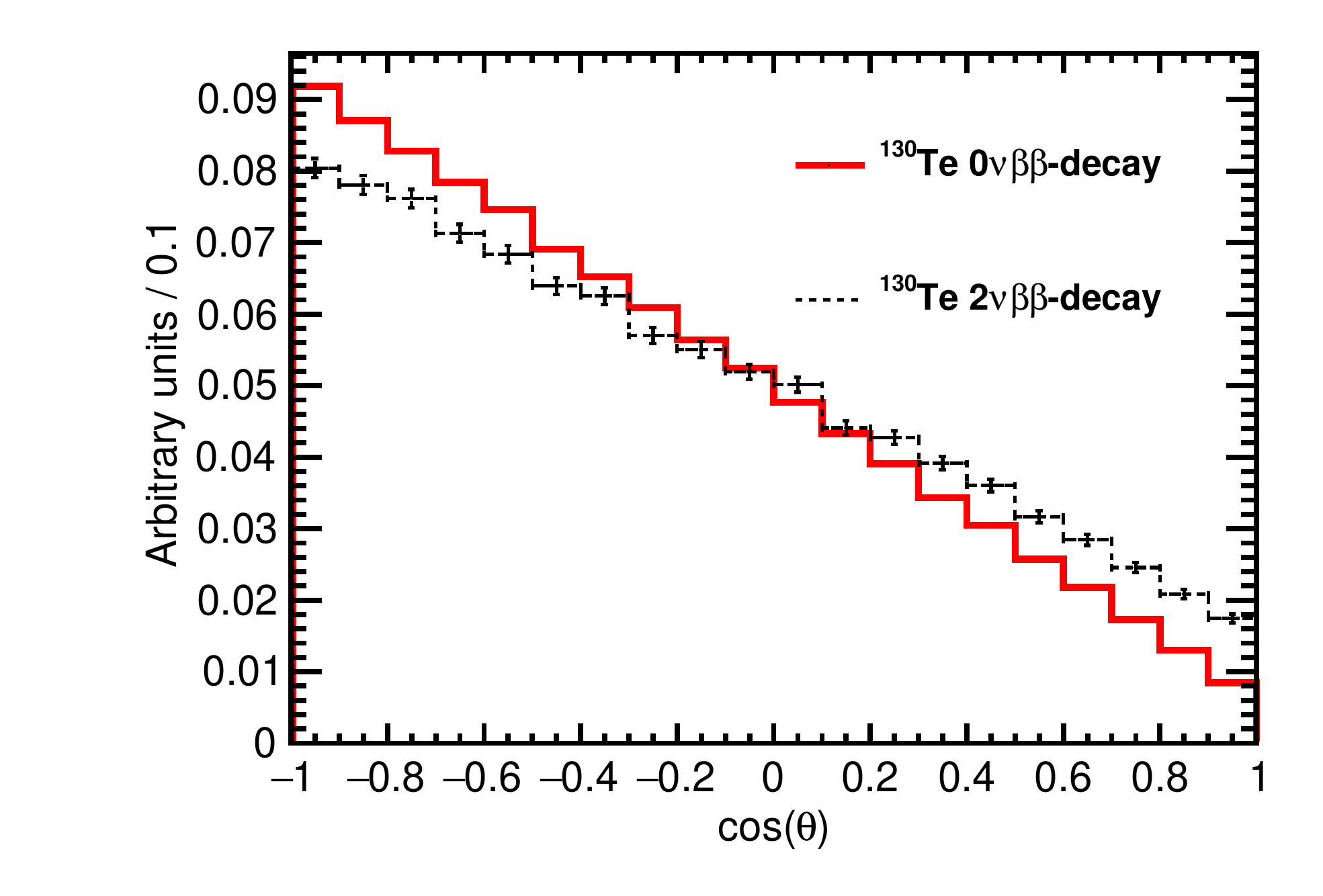}
  \includegraphics[width=0.49\textwidth]{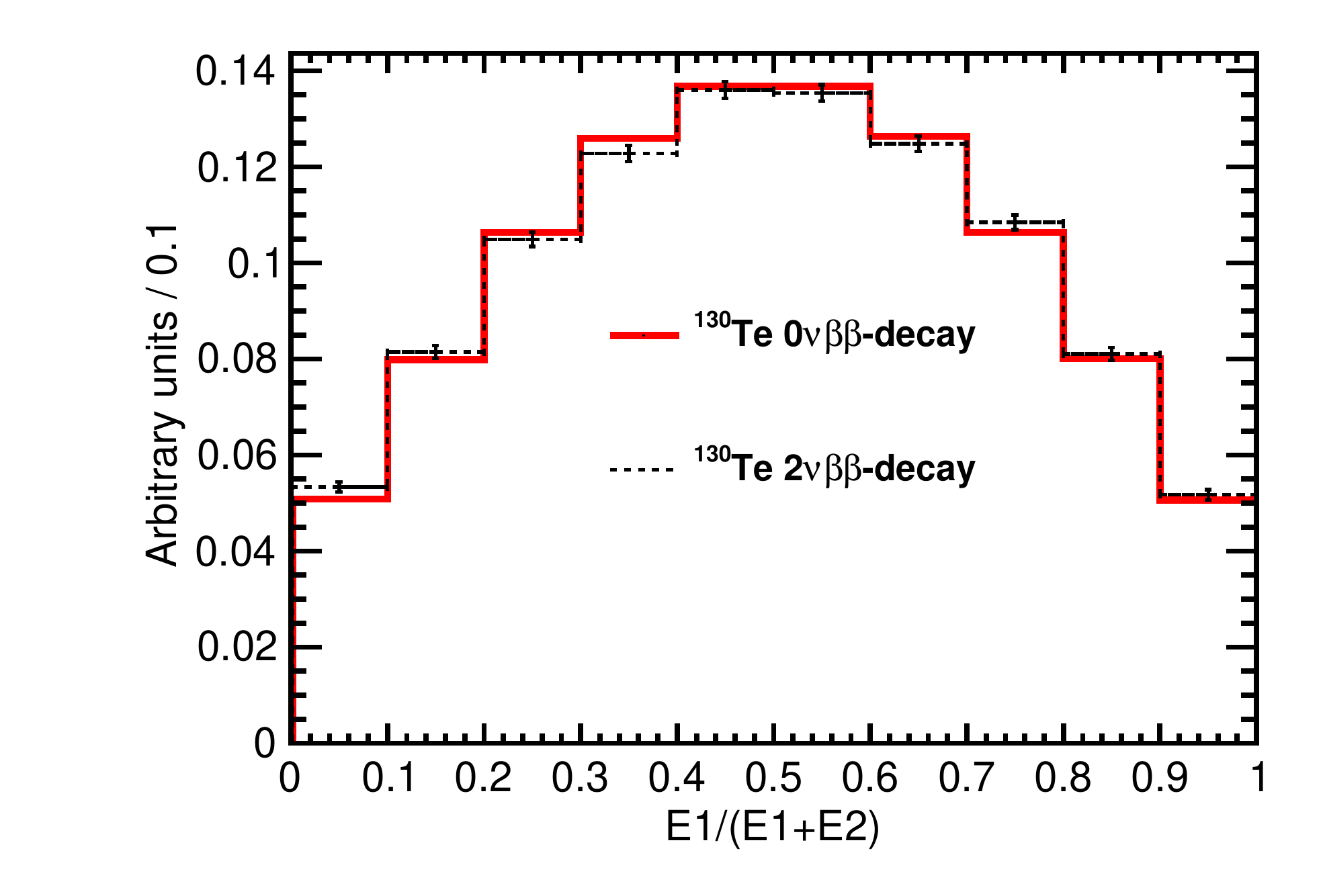}
  \caption{\emph{Left:} The distribution in the cosine of the angle
      between the two electrons for 0\nbb-decays (\emph{solid red line}).
      \emph{Right:} The fraction of the total energy carried by one of
      the two electrons in 0\nbb-decays (\emph{solid red line}).  In both
      panels the dashed black line is the corresponding distribution for
      SM 2\nbb-decay events with the total kinetic energy of the electrons
      above 95\% of the Q-value.}
  \label{fig:Kinematics}
\end{figure*}

The distribution in
$\cos{(\theta)}$ between the two electrons is presented in the left-hand
panel of Fig.~\ref{fig:Kinematics} (solid line), showing the
preference towards a back-to-back topology.  The energy sharing between
the electrons peaks at an equal split, as shown in the right-hand
panel of Fig.~\ref{fig:Kinematics} (solid line).

\subsection{Comparison to SM 2\nbb-decay}
\label{comparison}

Figure~\ref{fig:Kinematics} also shows the angular separation and
energy sharing of the two electrons in SM 2\nbb-decay events with the total kinetic
energy of the electrons above 95\% of the Q-value, found using the
same Monte Carlo generator but with SM phase space
factors~\cite{Jenni}.  As seen from the plot, the electron angular
correlations for 0\nbb-decay are slightly more back-to-back than those
from 2\nbb-decay due to a contribution from the neutrino
wave-functions even at vanishingly small energies of the
neutrinos~\cite{Jenni}. The energy sharing is essentially
identical.

\subsection{Production and Selection 
of Cherenkov light by electrons from \Te~ 0\nbb-decays}
Figure~\ref{fig:Energy_spectrum} also shows the threshold for the
production of Cherenkov light.
Examining the kinematics for one of the electrons from \Te~
0\nbb-decay with an equal energy split, the 1.26~MeV electron travels
on average a total path length of 7.1$\pm$0.9~mm, has a distance from
the origin of 5.6$\pm$1.0~mm in 26 $\pm$4~ps, and takes
24$\pm$3~ps to drop below Cherenkov threshold.  We note that due
to scattering of the electron, the final direction of the electron
before it stops does not match the initial direction; however,
the scattering angle is small at the time that the majority of
Cherenkov light is produced.

\begin{figure*}[ht]
  \centering
  \includegraphics[width=0.45\textwidth]{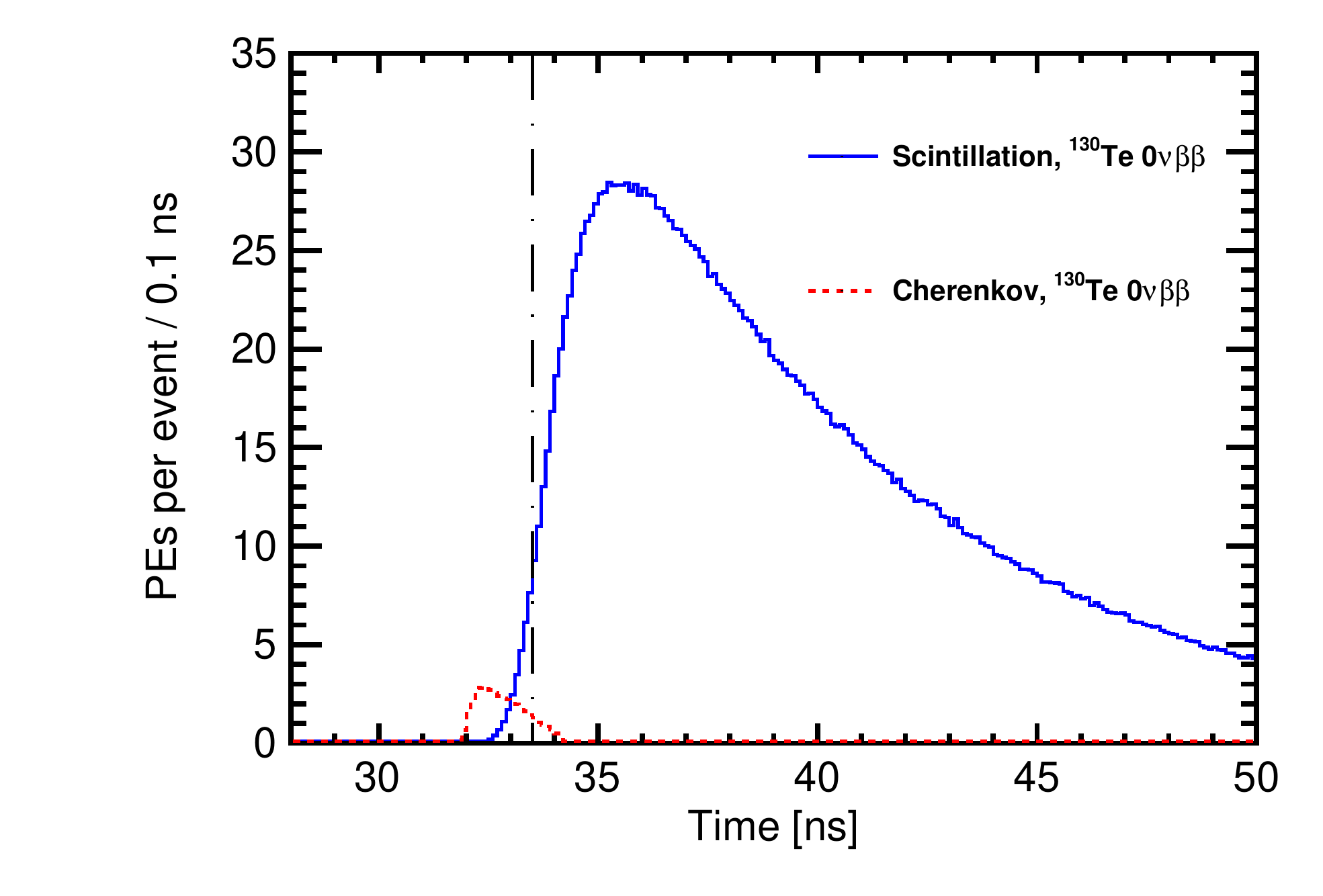}
\hfil
  \includegraphics[width=0.45\textwidth]{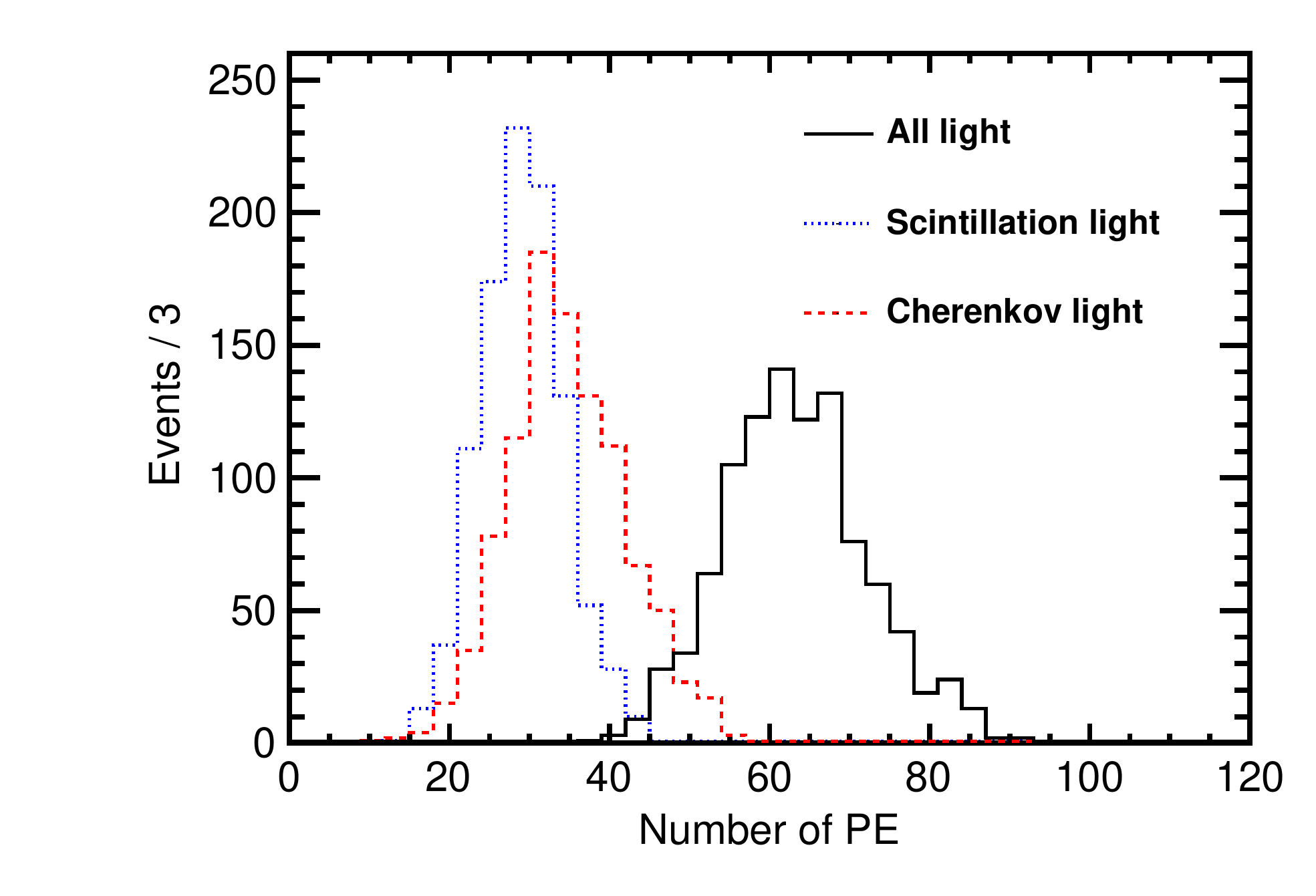}
  \caption{\emph{Left:} PE arrival times after
    application of the photo-detector TTS of
    100~ps for the default simulation of \Te~0\nbb-decay produced at
    the center of the detector.  Scintillation PEs (blue solid line)
    are compared to Cherenkov PEs (red dotted line). The vertical line
    at 33.5~ns indicates the time cut for the selection of the early
    PE sample.  \emph{Right:} Composition of the early PE sample (to
    the left of the vertical line in the left-hand panel):
    the number of Cherenkov (\emph{dashed red line}), scintillation
    (\emph{dotted blue line}), and total (\emph{solid black line}) PEs
    per event.}
\label{fig:ArrivalTimeDist}
\end{figure*}

Figure~\ref{fig:ArrivalTimeDist} shows distributions from the detector
simulation for 1000 \Te~ 0\nbb-decay events at the center of the
detector. The left-hand panel compares the time of PE
arrival at a photodetector anode for Cherenkov and scintillation
light, assuming a TTS in the
photodetector of 100 ps.  A selection of the PEs with relatively early
arrival time creates a sample with a high fraction of directional
Cherenkov light, designated as the `early PE' sample.

The right-hand panel shows the composition of the early PE sample,
selected with a time cut of 33.5~ns (vertical line on plot). On
average each \Te~ 0\nbb-decay produces 62.8$\pm$0.3 PEs in the early
PE sample, with an RMS width of 8.9 PEs from event-by-event
fluctuations.  On average the early PE sample consists of 28.6$\pm$0.2
scintillation PEs and 34.2$\pm$0.2 Cherenkov PEs, with RMS
distribution widths of 5.2 and 7.3 PEs respectively.

\subsection{\B~ solar neutrino background}

For a detector similar to our model, the \B~solar neutrino background is
significant due to the large total mass of the liquid scintillator in the
active region.  Electrons from elastic scattering of \B~solar
neutrinos have nearly a flat energy spectrum around the
Q-value~\cite{SNOp-B8-bkg}. We simulate \B~background as a single
monochromatic electron with energy of 2.53~MeV (Q-value of \Te). A
2.53~MeV electron travels a total path length of 15.5$\pm$2.0~mm, has
a distance from the origin of 12.6$\pm$2.2~mm in 55$\pm$7~ps, and
takes 49$\pm$2~ps to drop below Cherenkov threshold.

The shape of scintillation and Cherenkov PE timing distributions in
\B~events match very closely the shape of corresponding distributions
for 0\nbb-decay events shown in Fig.~\ref{fig:ArrivalTimeDist}. The
electron path length is too short compared to the detector size to
introduce any noticeable difference in the shape of PE timing
distributions between a single electron from \B~events and two
electrons from 0\nbb-events. 

\begin{figure*}[ht]

  \centering
  \includegraphics[width=0.45\textwidth]{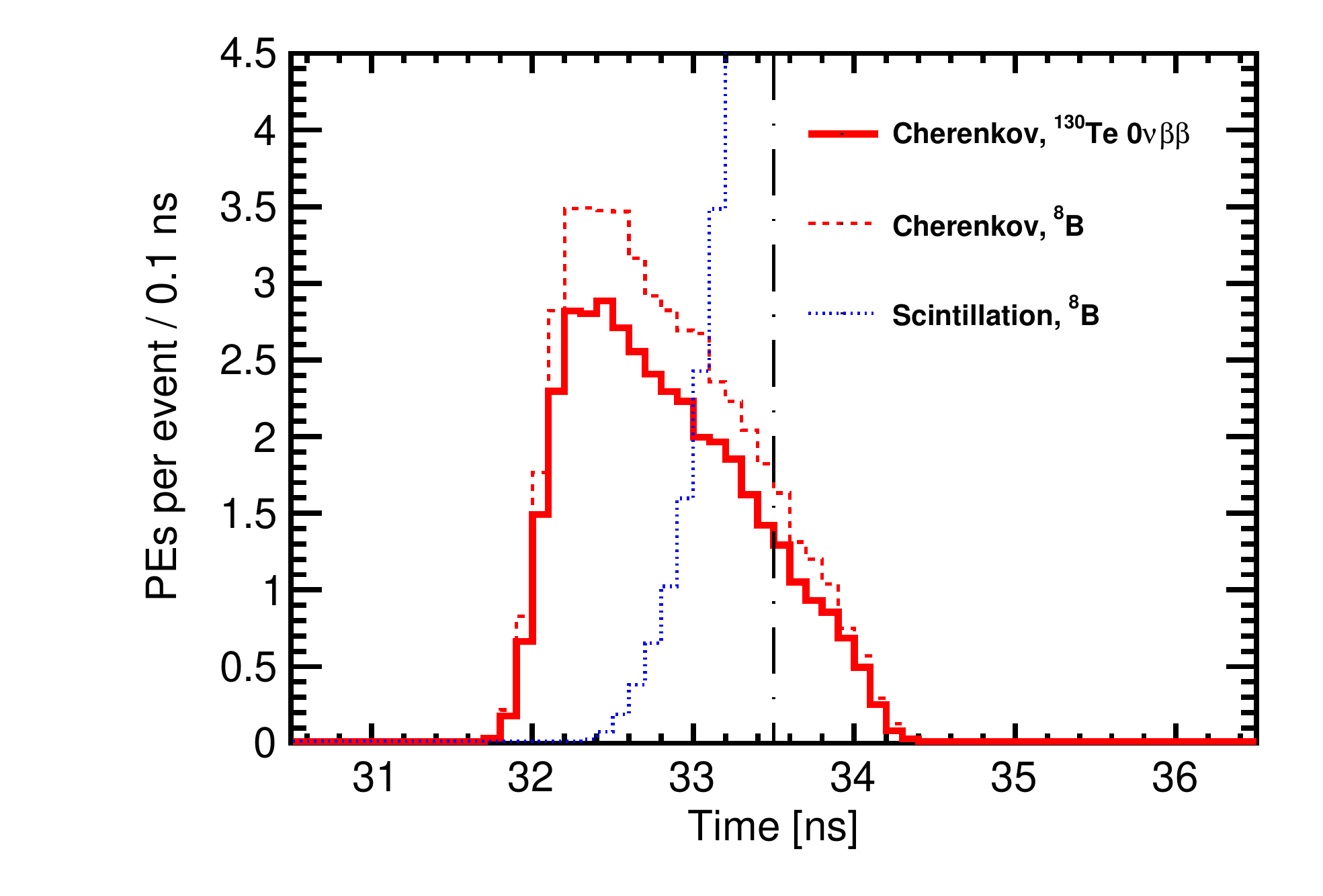}
\hfil
  \includegraphics[width=0.45\textwidth]{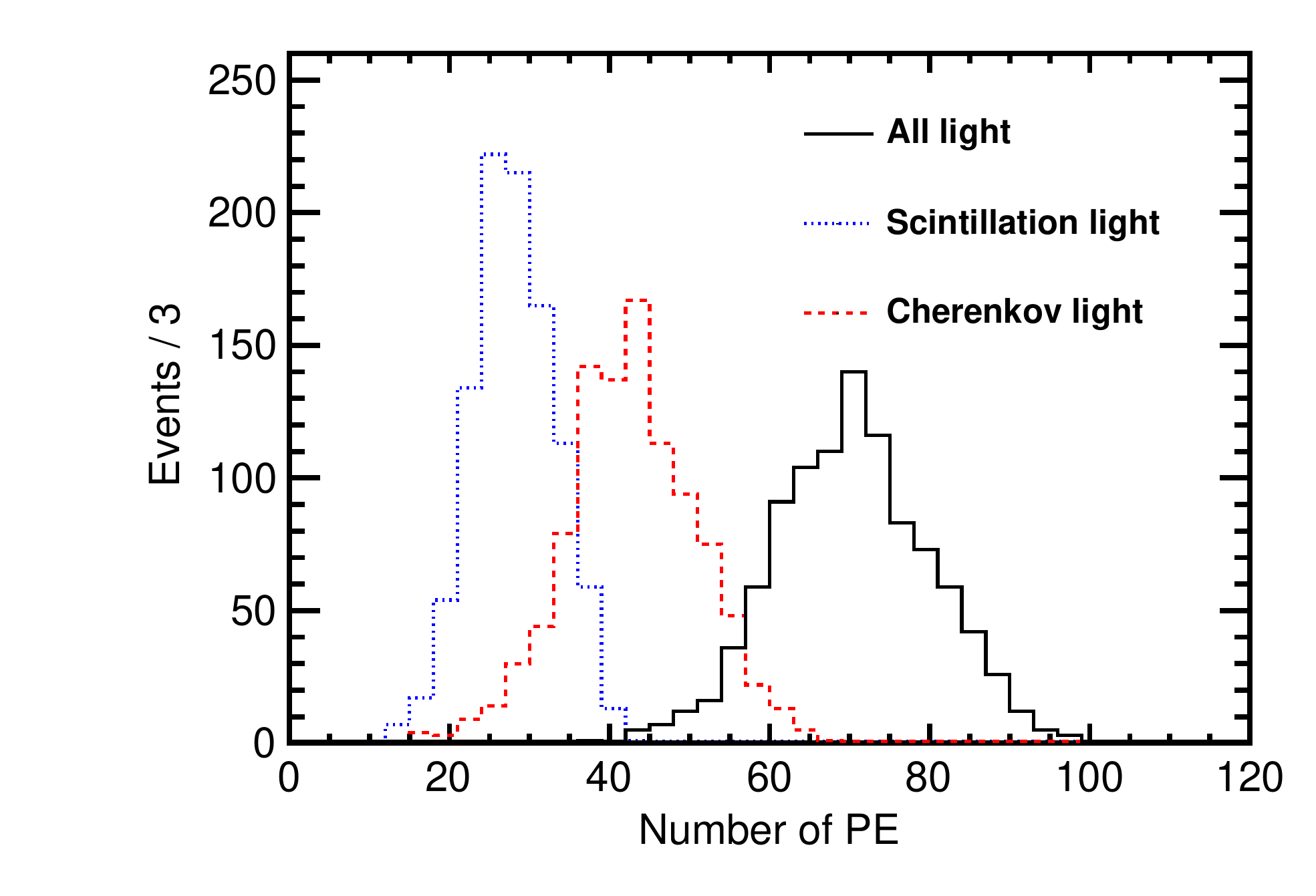}
  \caption{\emph{Left:} A `zoomed-in' view of the PEs
   arrival times of signal and background events produced at the
   center of the detector.  The distribution in time for Cherenkov
   PEs from \Te~0\nbb-decay is shown in solid black; Cherenkov PEs from
   \B~ solar neutrino background are shown in dashed red. PEs from
   scintillation are shown as the blue solid line. The line at 33.5~ns
   indicates the cut for the early PE sample selection.  \emph{Right:}
   Composition of the early PE sample: the number of Cherenkov PEs
   (\emph{dashed red line}), scintillation PEs (\emph{dotted blue
   line}), and total (\emph{solid black line}) PEs per event.}

\label{fig:ArrivalTimeDist_B8}
\end{figure*}

On average each \B~ neutrino event produces 69.9$\pm$0.3 PEs in the
early PE sample, with an RMS distribution width of 9.7 PEs due to
event-by-event fluctuations.  On average the early PE sample consist
of 27.6$\pm$0.2 scintillation and 42.3$\pm$0.3 Cherenkov PEs, with
event-by-event fluctuations contributing an RMS width of 5.2 and 8.2
PEs, respectively.  The total energy deposited in the detector in \B~
solar neutrino and 0\nbb-decay events is the same. This leads to
nearly the same amount of scintillation light produced in the
detector.

The number of Cherenkov photons is $\sim$10\% higher for \B~neutrino
events compared to 0\nbb-decay events. This is because Cherenkov light
in \B~ neutrino interactions is being produced by a single electron,
while the same kinetic energy is split between two electrons in
0\nbb-decay events~\footnote{We do not use the small difference in the
total number of PEs in the early PE sample due to the Cherenkov PE
contribution to separate 0\nbb-decay signal from \B~background.
However, it may provide an extra handle on signal-background
separation in a multivariate analysis when combined with directional
and topographical information.}.

\section{Event Topology and the Spherical Harmonics Analysis}
\label{sec:topology_and_harmonics}

We have developed a method based on a spherical harmonics
decomposition to discriminate the topologies of 0\nbb-decay
two-electron events and \B-neutrino single-electron events. The
identification of the Cherenkov photon clusters is challenging due to
the smearing of the characteristic ring pattern by multiple scattering
of the electrons and by the smallness of the Cherenkov signal
relative to the large amount of uniformly-distributed scintillation
light.  We find that performing the spherical harmonics analysis on
the smaller early PE sample, which has a relatively high fraction
of Cherenkov PEs, can discriminate 0\nbb-decay signal events from
backgrounds, although a high rejection factor will require a slower
scintillator than in the model.

\subsection{Topology of 0\nbb-decay and \B~Events}
\label{subsec:topology}

With \Te~as the active isotope, all background from \B~solar neutrinos
will have the single electron above Cherenkov threshold in the liquid
scintillator. Also, a large fraction of 0\nbb-decay signal events will
have both electrons above Cherenkov threshold.

In some cases only one Cherenkov cluster is produced in 0\nbb-decay
signal events. This happens either when the angle between the two
0\nbb-decay electrons is small and Cherenkov clusters overlap or when
the energy split between electrons is not balanced, causing one
electron to be below Cherenkov threshold.  Such signal events cannot
be separated from background based on the topology of the distribution
of Cherenkov photons on the detector surface.  However, the
directionality of the electron that is above Cherenkov threshold can
still be reconstructed. This directionality information may allow for
suppression of \B~events based on the position of the
sun~\cite{theia_white_paper}.

\begin{figure*}[ht]
  \centering
  \includegraphics[width=0.4\textwidth]{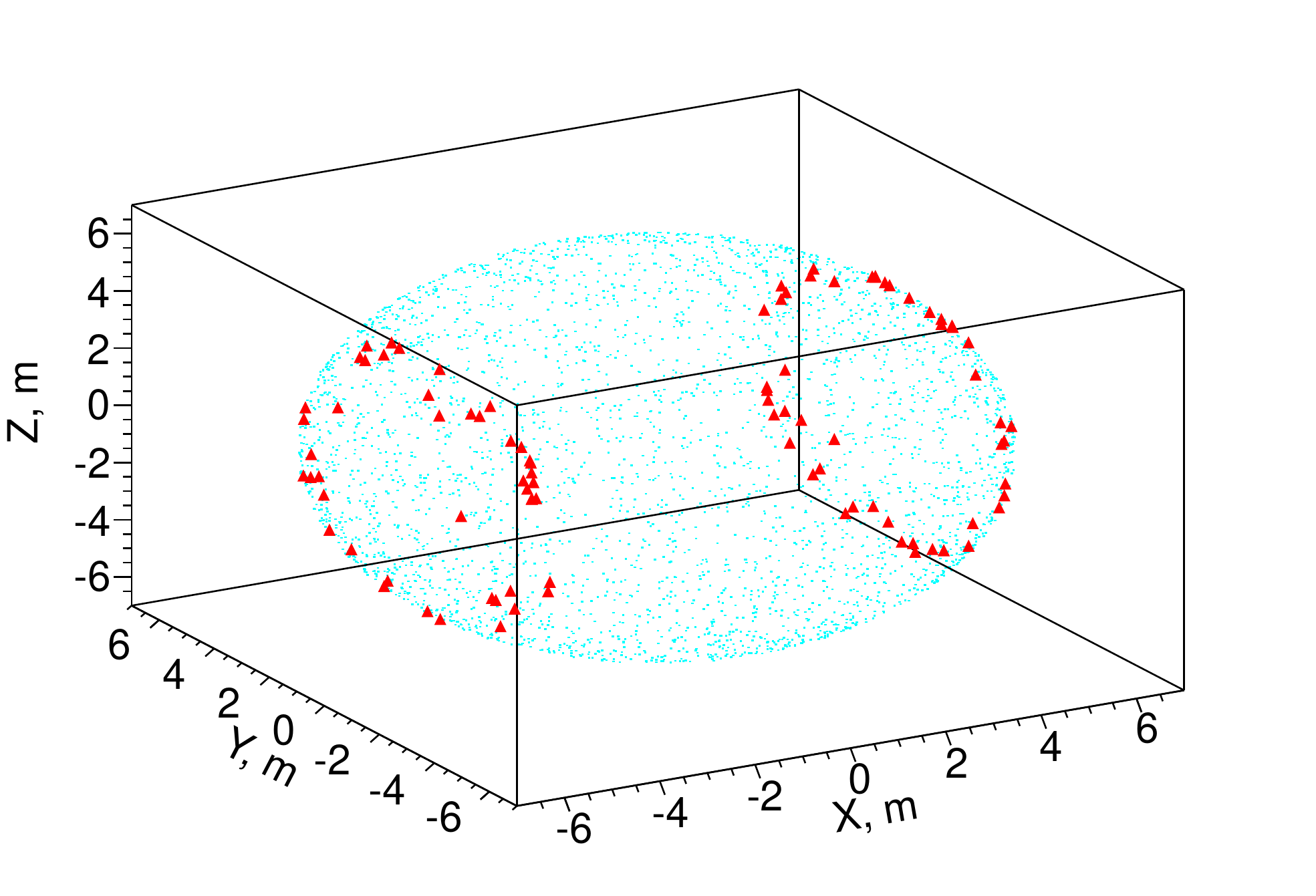}
  \includegraphics[width=0.4\textwidth]{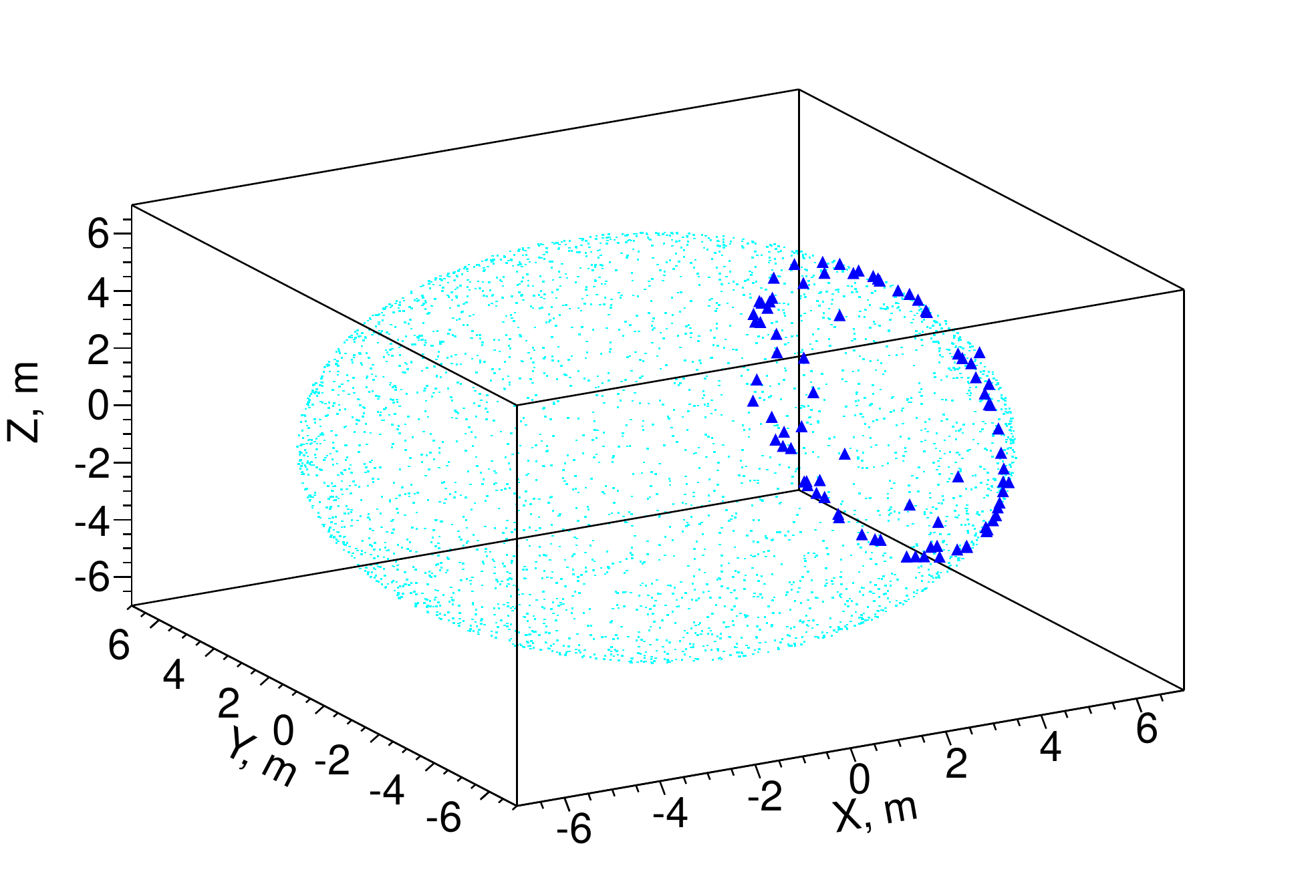}
  \includegraphics[width=0.7\textwidth]{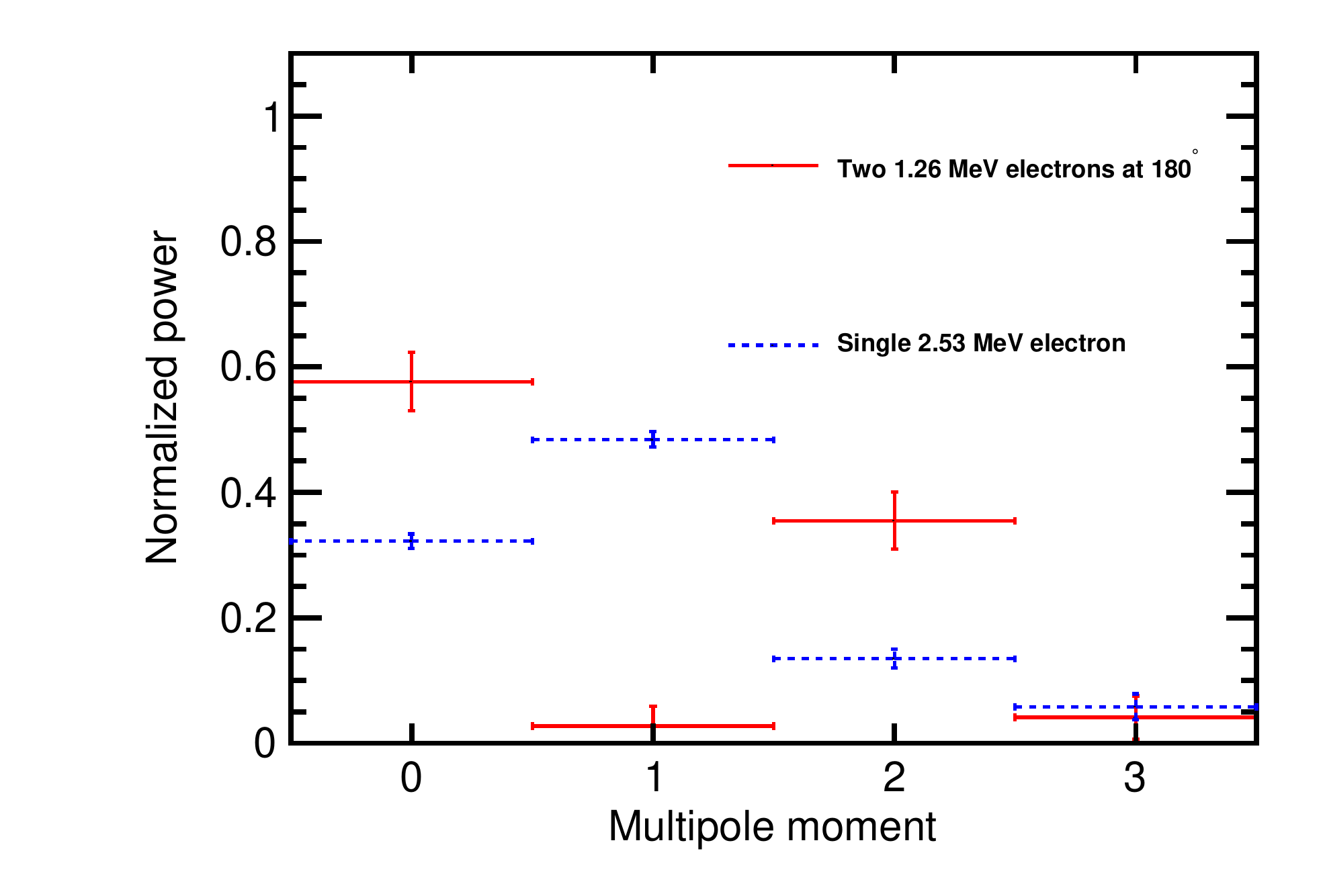}
  \caption{\emph{Top panels:} Idealized event displays, with multiple
scattering turned off and at the center of the detector, of:
(\emph{top left}) a signal event with two 1.26~MeV back-to-back
electrons; and (\emph{top right}) a \B~neutrino background event with
single 2.53~MeV electron. A 30\% QE is assumed for both Cherenkov
PEs (\emph{triangles}) and scintillation PEs (\emph{dots}).  
\emph{Bottom panel:} The normalized power spectrum $S_{\ell}$ for the
Cherenkov PEs only, calculated event-by-event for 100 events for the two
above topologies. The heights of the vertical bars correspond to event by event
variation $(\pm 1 ~\sigma)$.}
  \label{fig:ThreeTopologies_Display_NoMultScat}
\end{figure*}

For the purpose of illustrating the spherical harmonics analysis
concept, we first consider two distinct topologies: a) two electrons
produced back-to-back at an 180$^{\circ}$ angle;  and b) a single electron.
Figure~\ref{fig:ThreeTopologies_Display_NoMultScat} shows an idealized
simulation of these two topologies for a total electron energy of
2.53~MeV. In order to emphasize ring patterns formed by Cherenkov
photons, the electron multiple scattering process is turned off in this
idealized simulation and a photocathode QE of 30\% is used for both
Cherenkov and scintillation photons. Here the single-electron event
represents an idealized \B~event topology and the two-electron events
represent two special cases of an idealized 0\nbb-decay topology.

\subsection{Description of the Spherical Harmonics Analysis}

The central strategy of the spherical harmonics analysis is to
construct rotationally invariant variables that can be used to
separate different event topologies. To account for the fluctuation of
the number of PEs from event to event, we use a normalized power,
$S_{\ell}$, defined in Appendix A.

The bottom panel in Fig.~\ref{fig:ThreeTopologies_Display_NoMultScat}
compares the normalized power spectra for the two representative event
topologies in the idealized case of no multiple scattering and with a 30\%
quantum efficiency for both Cherenkov and scintillation
photons. In this case, the method gives a good separation between the
two event topologies.

\begin{figure*}[ht]
  \centering
  \includegraphics[width=0.4\textwidth]{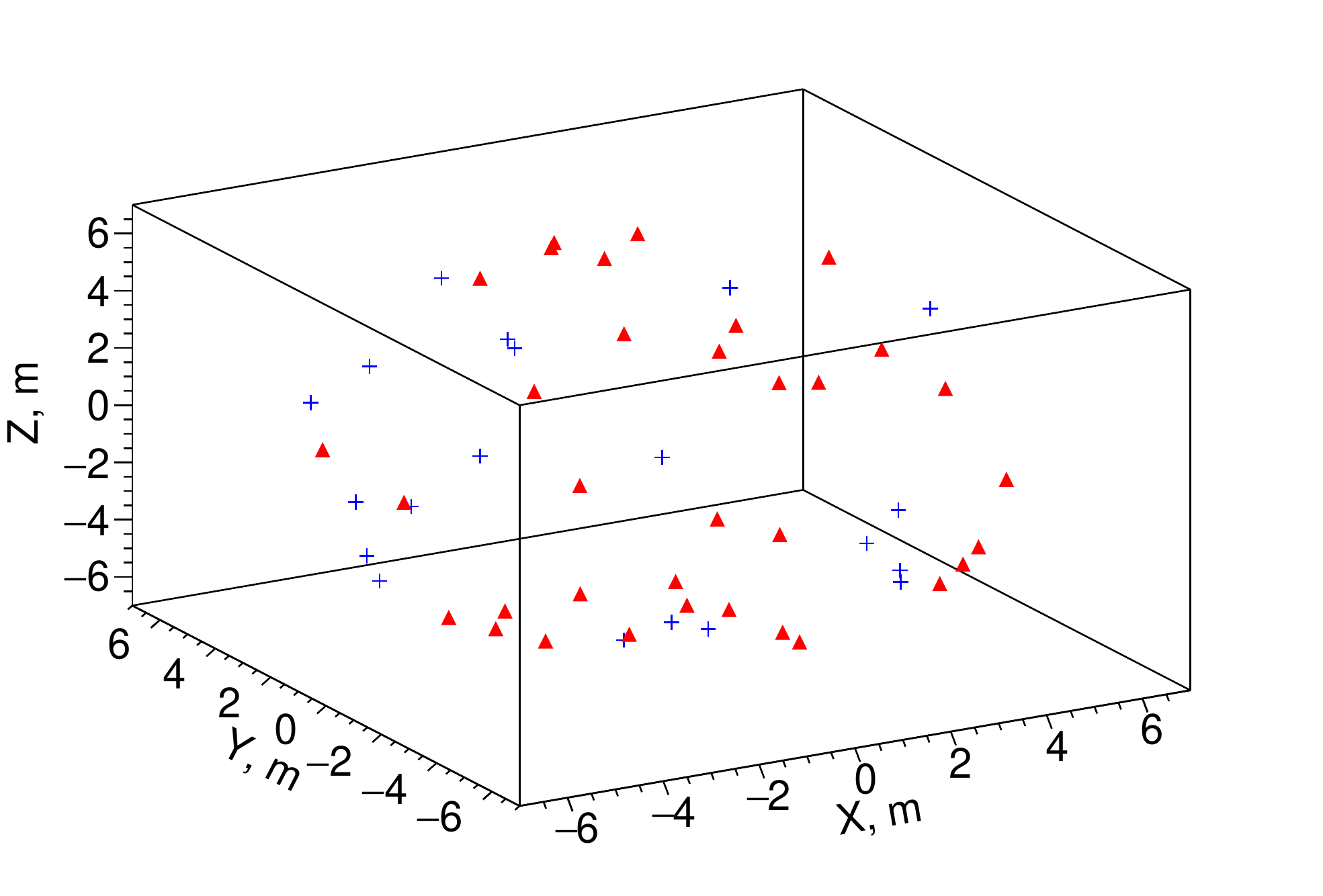}
  \includegraphics[width=0.4\textwidth]{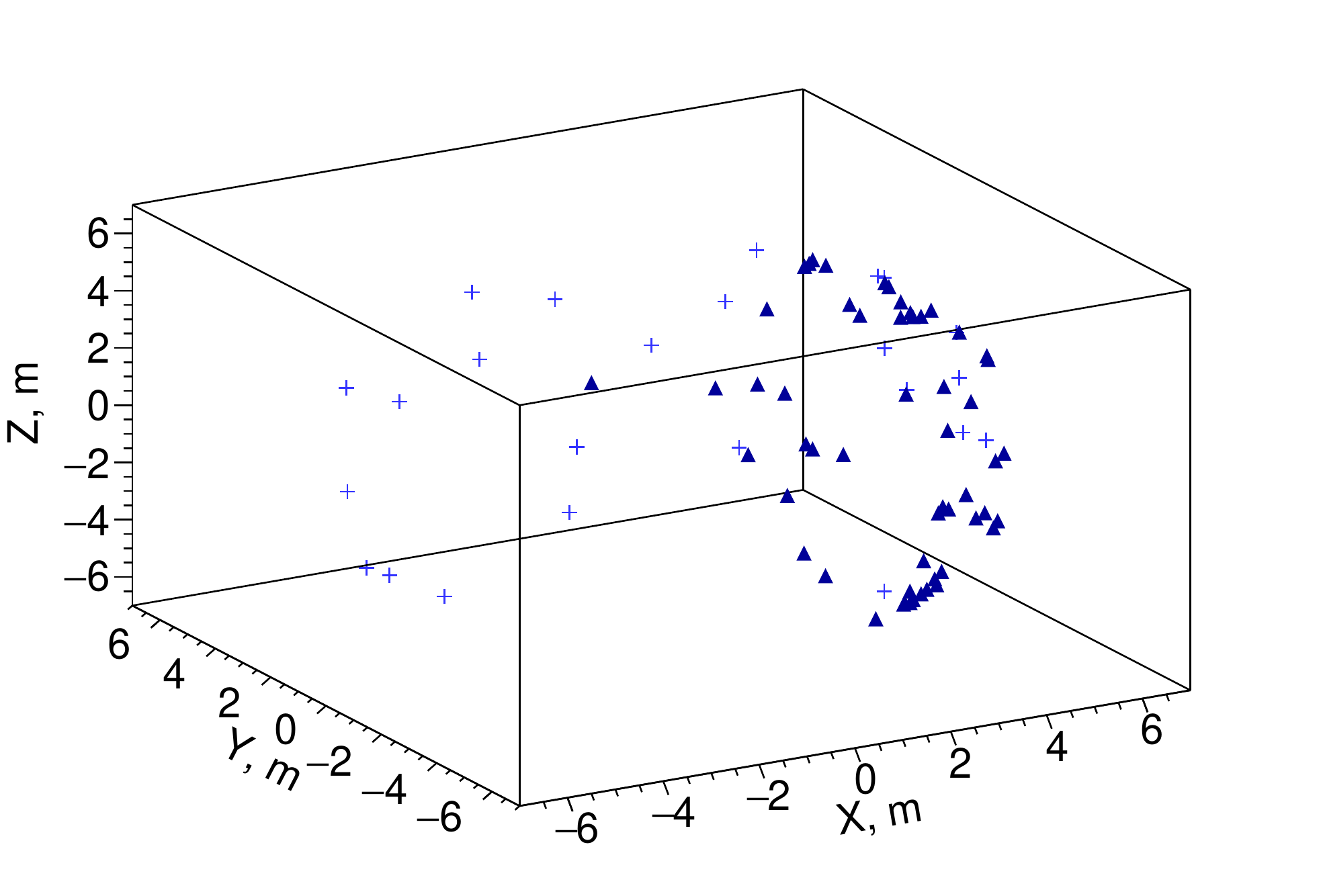}
  \includegraphics[width=0.7\textwidth]{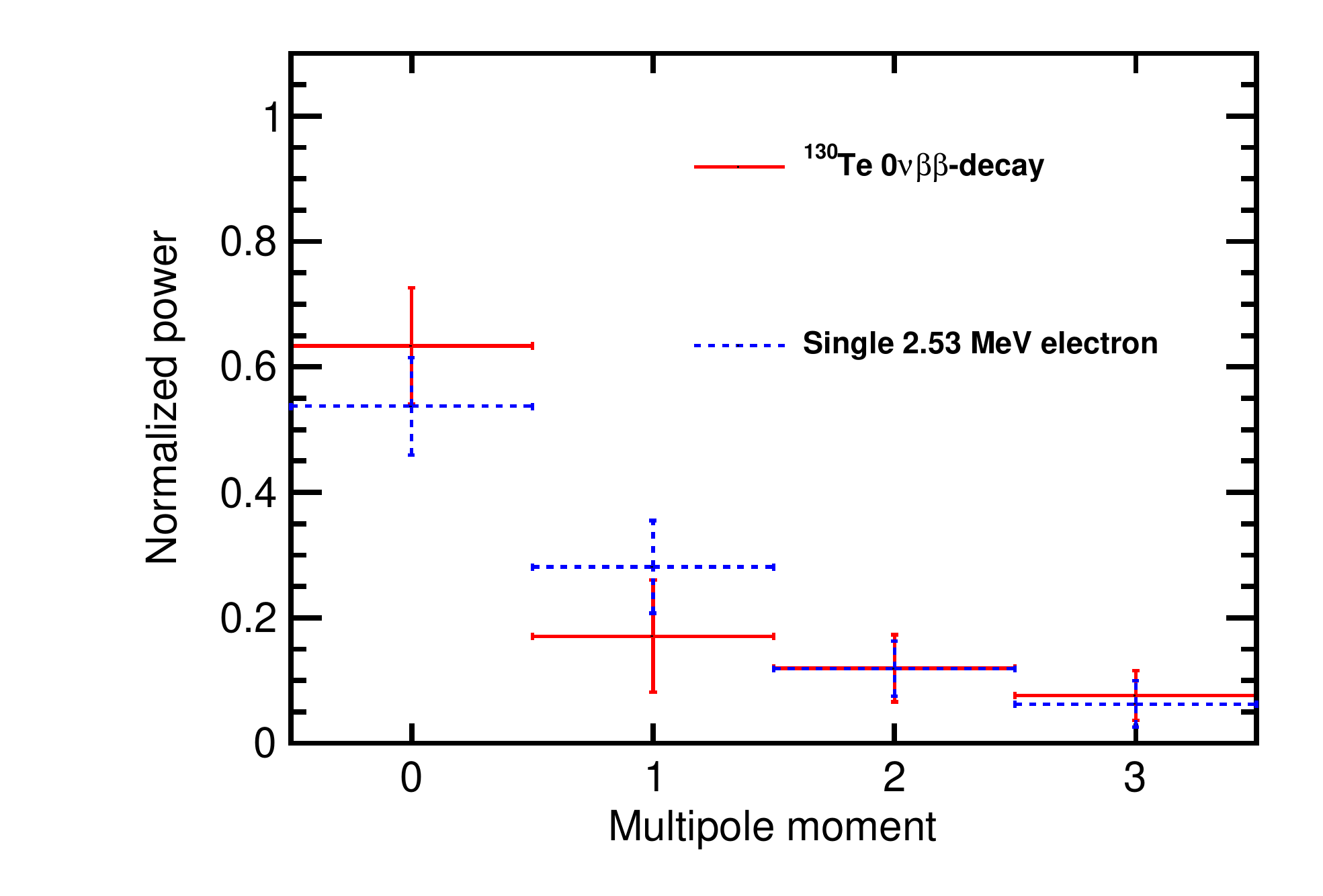} 
  \caption{\emph{Top panels:} Event displays with multiple scattering
and at the center of the detector for: (\emph{top left}) a signal
event with two 1.26~MeV back-to-back electrons; and (\emph{top right}) 
a \B~neutrino background event with a single 2.53~MeV electron. Only early
PEs are shown. The
model QEs are assumed for both Cherenkov PEs (\emph{triangles}) and
scintillation PEs (\emph{crosses}). \emph{Bottom panel:} The normalized
power spectrum $S_{\ell}$ for all early PEs, calculated
event-by-event for 1000 events of 0\nbb-decay signal and \B~neutrino
events. The
heights of the vertical bars corespond to event by event
variation $(\pm 1 ~\sigma)$.}
\label{fig:Te130_Display}
\end{figure*}

At energies relevant to 0\nbb-decay, multiple scattering makes the Cherenkov 
rings fuzzy. In most cases,
$\sim$1~MeV electrons produce randomly shaped clusters of Cherenkov
photons around the direction of the electron track.  Examples of \Te~
0\nbb-decay and \B~events simulated with multiple scattering, but still at
the center of the detector, are shown in Fig.~\ref{fig:Te130_Display}.
\Te~ events are generated based on the phase factors described
in~\cite{Jenni}.  $^{8}$B events are implemented as
monochromatic electrons with the initial direction along the
$x$-axis. The default QEs of 12\% for Cherenkov light and 23\% for
scintillation light have been  applied. Figure~\ref{fig:Te130_Display} shows
early PEs that pass the 33.5~ns time cut. 

In this more realistic example, the uniformly distributed
scintillation light makes it difficult to visually distinguish
the event topology. The power spectra shown in the bottom panel of
Fig.~\ref{fig:Te130_Display}~ are different only at $\ell$=0 and
$\ell$=1. We use this difference to separate 0\nbb-decay
signal from \B~background events.

As expected, we find that 0\nbb-decay events become indistinguishable from single-track events when
the angle between the two electrons is small and two Cherenkov
clusters overlap. Event topologies of 0\nbb-decay and \B~events are also
very similar when only one electron from 0\nbb-decay is above the Cherenkov
threshold. The spherical harmonics analysis is most efficient
for events with large angular separation between the two electrons and
when both electrons are above Cherenkov threshold~\cite{further_cuts}.

\section{Performance of the Spherical Harmonics Analysis in Separating 0\nbb-decay from \B~Background.}
\label{sec:performance}

The separation of signal and background comes almost entirely from the
first two multipole moments, $\ell$=0 and $\ell$=0. However, higher multipole
moments are needed for the event-by-event normalization of the power spectrum
$S_{\ell}$ (Eq.~\ref{eq6}). In the following, we choose to calculate the
power spectrum $s_{\ell}$ up to $\ell$=3 and use only the normalized variables $S_0$
and $S_1$, where the normalization is given by

\begin{eqnarray}
\label{eq7}
S_{0,1} = \frac{s_{0,1}}{\sum_{l=0}^{3} s_{\ell}}.
\end{eqnarray}

As discussed below, a linear combination of $S_0$ and $S_1$ can be
used to separate 0\nbb-decay and \B~events.

\subsection{Central events with no uncertainty on the vertex position}

To illustrate the technique, we initially evaluate 
the performance of the spherical harmonics
analysis in the idealized case of events at the center of the
detector with perfect reconstruction of the event vertex
position. 

\begin{figure*}[h]
\centering
\includegraphics[width=0.49\textwidth]{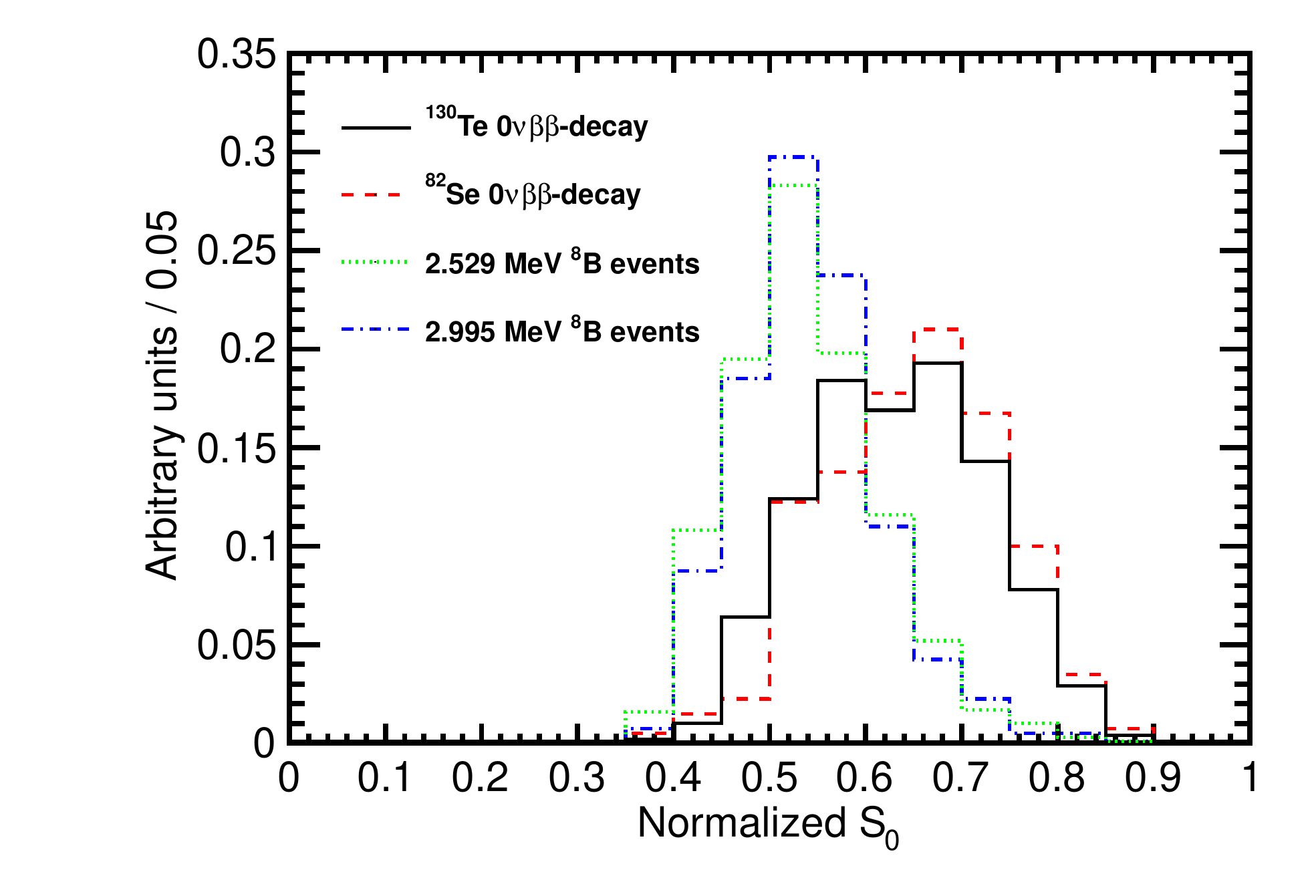}
\includegraphics[width=0.49\textwidth]{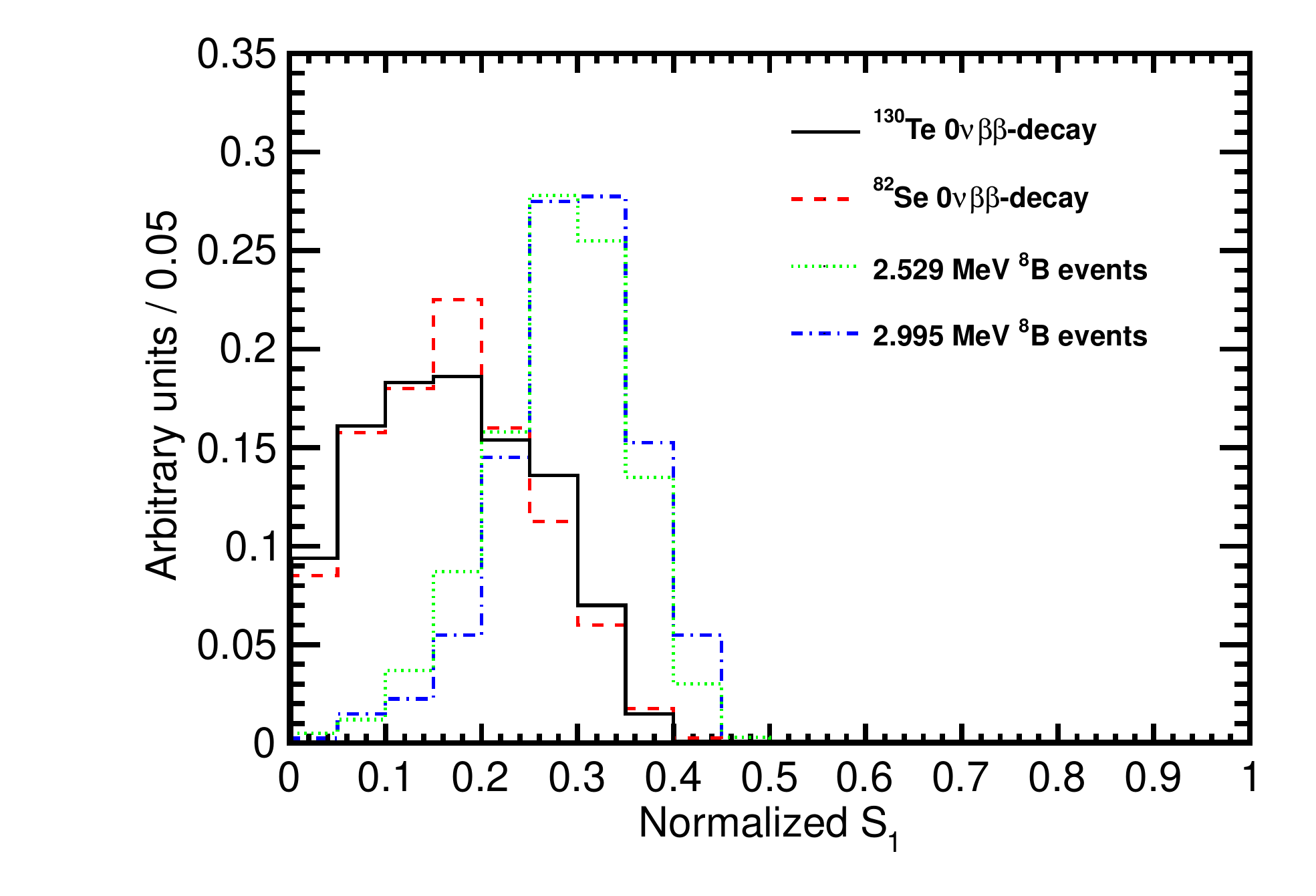}
\caption{Results from the idealized case of central events at the
  detector origin( i.e. perfect vertex reconstruction); a time cut of 33.5~ns
  on the PE arrival time is applied. The default QE and 100\%
  photo-coverage are used in the simulation.  (\emph{Left}) $S_0$ and
  $S_1$ (\emph{right}) distributions for 1000 simulated 0\nbb-decay
  signal and \B~background events.  Two different isotopes are
  compared, $^{130}$Te and $^{82}$Se. The corresponding kinetic
  energies of background \B~neutrino single electrons are 2.53 MeV and
  3.00 MeV.}
\label{fig:S_vs_energy}
\end{figure*}

Comparisons of $S_0$ and $S_1$ distributions for $^{82}$Se and $^{130}$Te 
0\nbb-decay signal and corresponding \B~background events are shown in 
Fig.~\ref{fig:S_vs_energy}. Both variables, $S_0$ and $S_1$, provide 
a noticeable separation between signal and background. 
$^{82}$Se 0\nbb-decay events are shown to demonstrate that in the energy 
range of interest, the $S_{\ell}$ do not strongly depend on the energy 
deposited in the detector, i.e. information contained in the normalized power
spectrum is complimentary to the energy measurements. 

\begin{figure*}[h]
  \centering
  \includegraphics[width=0.49\textwidth]{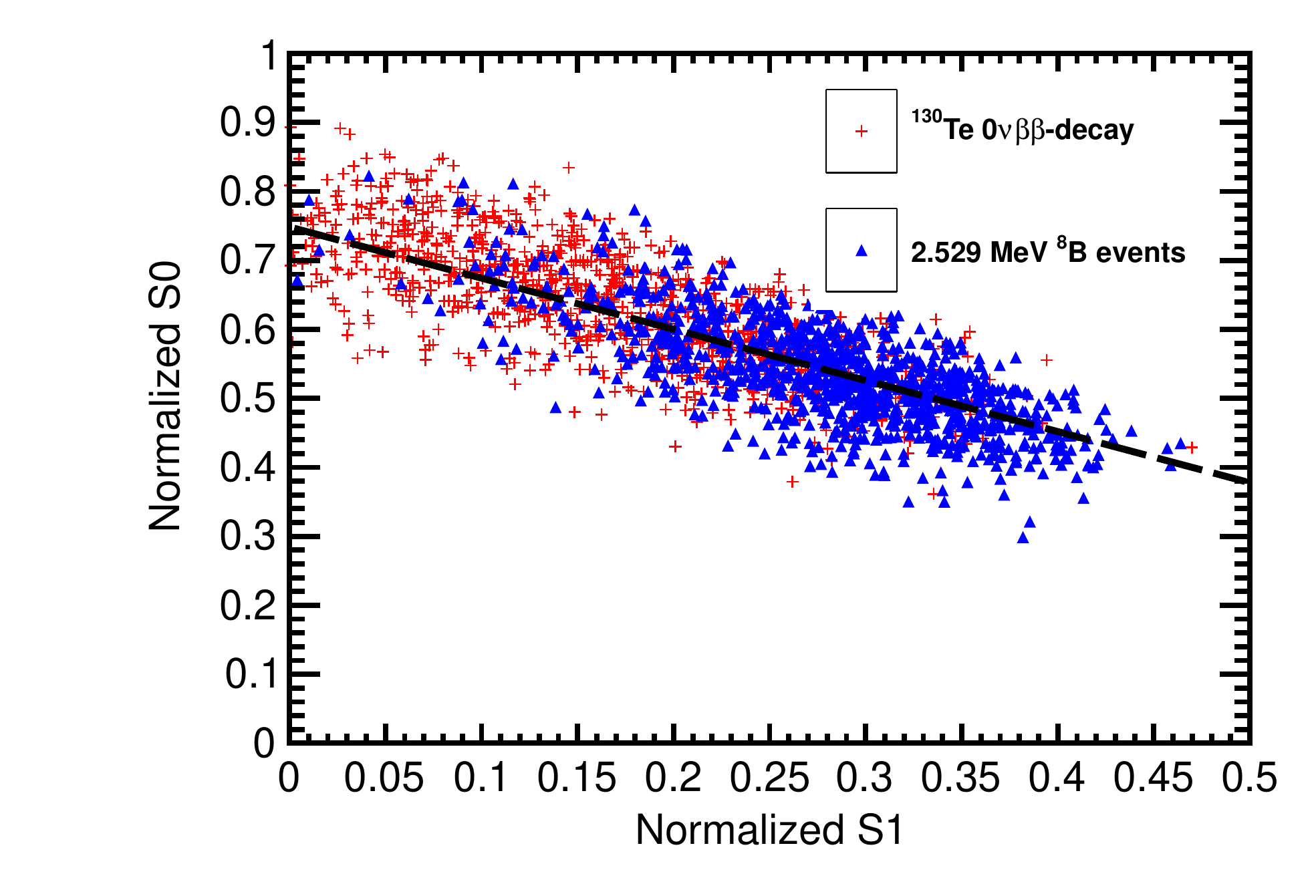}
  \includegraphics[width=0.49\textwidth]{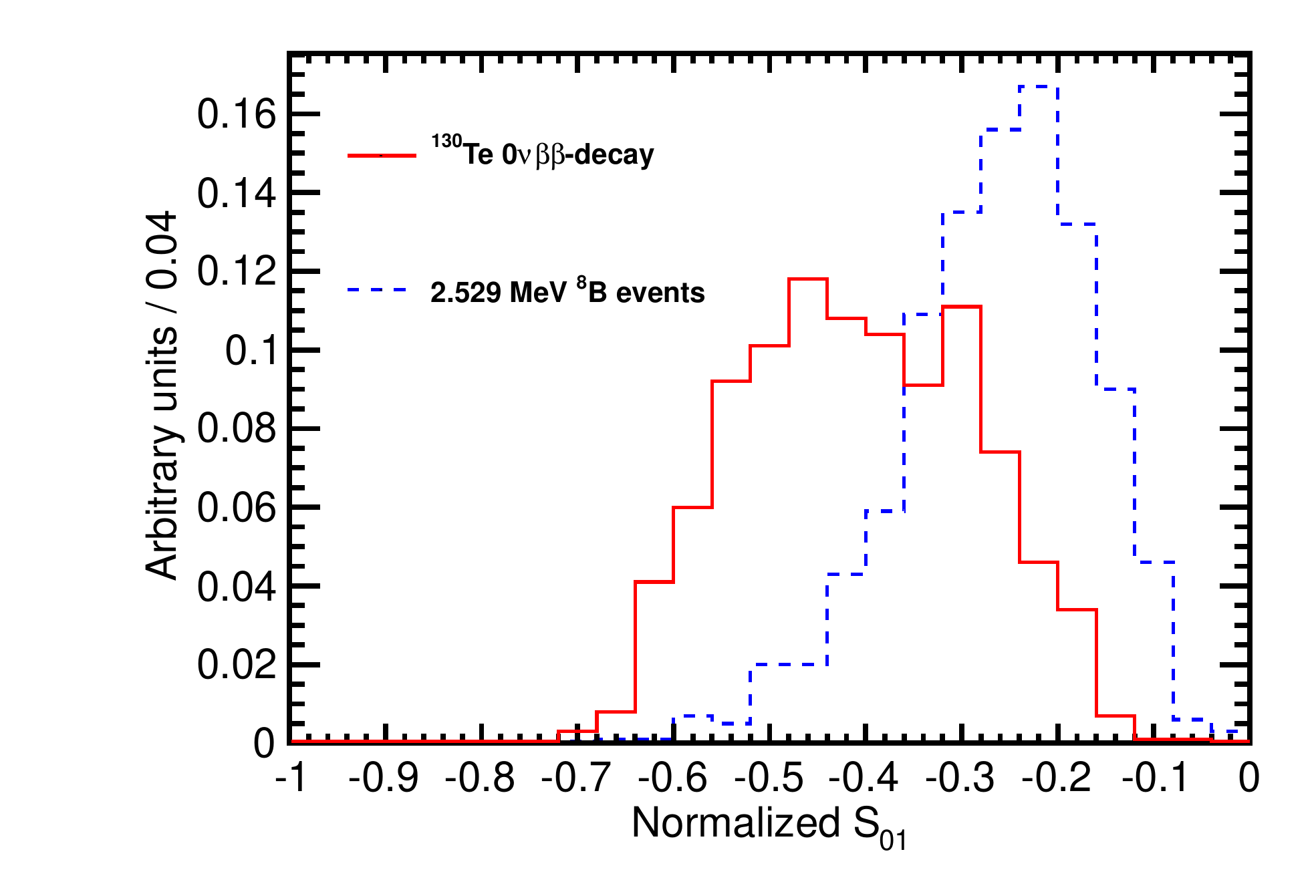}
  \caption{\emph{Left:} Scatter plot of the moments $S_0$ versus $S_1$
    for a simulation of 1000 signal (\emph{red crosses}) and
    background (\emph{blue triangles}), for the idealized case of
    central events assuming perfect reconstruction of the vertex
    position. A time cut of 33.5~ns on the PE arrival time is
    applied. The default QE and 100\% photo-coverage is used in the
    simulation.  The black dashed line corresponds to a linear fit for
    $S_{01}$.  \emph{Right:} Comparison of the $S_{01}$ distribution
    between signal (\emph{red solid line}) and background (\emph{blue
    dashed line}).  $I_{overlap}$=0.52.}
\label{fig:SL_Te_33p5ns_center}
\end{figure*}

The left-hand panel in Fig.~\ref{fig:SL_Te_33p5ns_center} compares
scatter plots of the first two components of the power spectrum, $S_0$
and $S_1$, for signal and background. In order to illustrate the
separation between $^{130}$Te and $^{8}$B events, a linear combination
of variables $S_0$ and $S_1$ is constructed as follows~\footnote{A
multi-variate event-by-event analysis will have more discriminatory
power than this simple 1-dimensional separation, but in the absence of
a real detector is a waste of time~\cite{goldberger_watson}.}.

First, we perform a linear fit to $S_0$ = $A \cdot S_1 + B$, of all points on the
scatter plot, as shown by the dashed line in the left-hand
panel in Fig.~\ref{fig:SL_Te_33p5ns_center}. A 1-dimensional (1-D) variable
$S_{01}$ is defined as $S_{01} = S_1 \cdot \cos(\theta) + S_0 \cdot
\sin(\theta)$, where $\tan(\theta)$=$A$. The right-hand panel in
Fig.~\ref{fig:SL_Te_33p5ns_center} compares distributions of $S_{01}$
for 0\nbb-decay signal and \B~background. These 1-D histograms for
$S_{01}$ represent the projection of the points on the scatter plot
onto the fitted line.

To quantify the separation between the signal and background we
calculate the area of the overlap in the $S_{01}$ distributions,
$I_{overlap}$. There is no separation if $I_{overlap}$=1, and there is
a 100\% separation if
$I_{overlap}$=0. Figure~\ref{fig:SL_Te_33p5ns_center} shows the
separation of this simple algorithm based on the shape of the early PE
sample; the overlap between signal and background is
$I_{overlap}$=0.52. At an efficiency for the signal of
70\% we find a rejection factor of 4.6.

\subsection{Events in a fiducial volume with an uncertainty on the vertex position}

We find that in the default detector model the separation power of
the spherical harmonics analysis is significantly reduced when event
vertex is not at the center of the detector and vertex resolution
is taken into account.

For the general case, even
significantly delayed scintillation photons can reach the side of the
detector that is closer to the vertex much earlier than Cherenkov
photons traveling to the opposite side of the detector. The time cut
thus has to take into account the total distance traveled by each
individual photon. In order to select early PE sample we use a 
differential cut of 
$\Delta t=t^{phot}_{measured} - t^{phot}_{predicted}<$1~ns, where 
$t^{phot}_{measured}$ is the measured time of the photon hit and
$t^{phot}_{predicted}$ is the predicted time based on the reconstructed 
vertex position.\footnote{ $t^{phot}_{predicted} = L^{phot}/v^{phot}$, 
where the $L^{phot}$ is the distance from the vertex to the photon hit 
on the detector sphere and $v^{phot}$ is the photon group velocity. 
Chromatic dispersion thus reduces the efficiency of the time cut in 
selecting early PE sample with high fraction of Cherenkov PE.}

In general, the $S_1$ component of the spherical harmonics power
spectrum is higher for asymmetric distributions and lower for
symmetric distributions (e.g., compare the back-to-back and single
electron topologies in
Fig.~\ref{fig:ThreeTopologies_Display_NoMultScat}). If a vertex is 
shifted in the direction opposite to the track of the electron, the
differential time cut selects more scintillation photons that are
emitted in the direction of the electron track.  Scintillation photons
would enhance the forward asymmetry of the early PE sample, which
in turn would move $S_1$ to higher values.  Moreover, $S_1=$~0 for a
distribution with perfect symmetry with respect to the center of the
sphere.  If a vertex is shifted in the same direction as the direction
of the electron, the differential time cut selects more scintillation
photons that are emitted in the direction opposite to the electron
track. The asymmetry of Cherenkov PEs would then be counter-balanced
by scintillation PEs, which in turn, would move $S_1$ to lower values.

We simulated 1000 signal and background events that have their
vertices uniformly distributed within a fiducial volume of $R<3$~m,
where $R$ is the distance between the event vertex and the center of
the detector, with a vertex resolution of 5.2~cm based on our earlier study of
reconstruction\cite{Aberle2014}. 
The uncertainty on the vertex reconstruction is implemented as smearing
along $x$, $y$, and $z$ directions with three independent Gaussian distributions 
of the same width, $\sigma_x = \sigma_y = \sigma_z =$3~cm.

\begin{figure*}[h]
  \centering
  \includegraphics[width=0.49\textwidth]{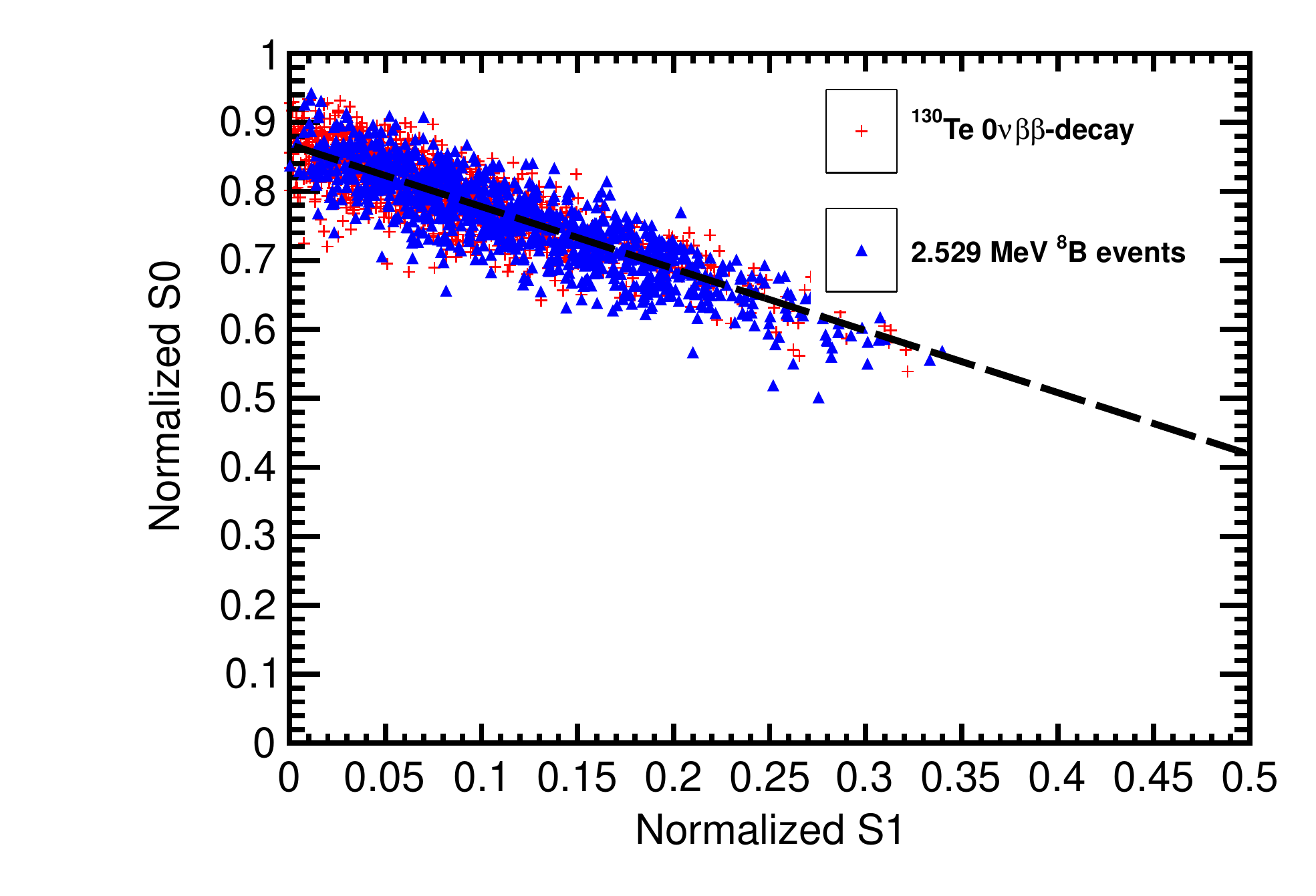}
  \includegraphics[width=0.49\textwidth]{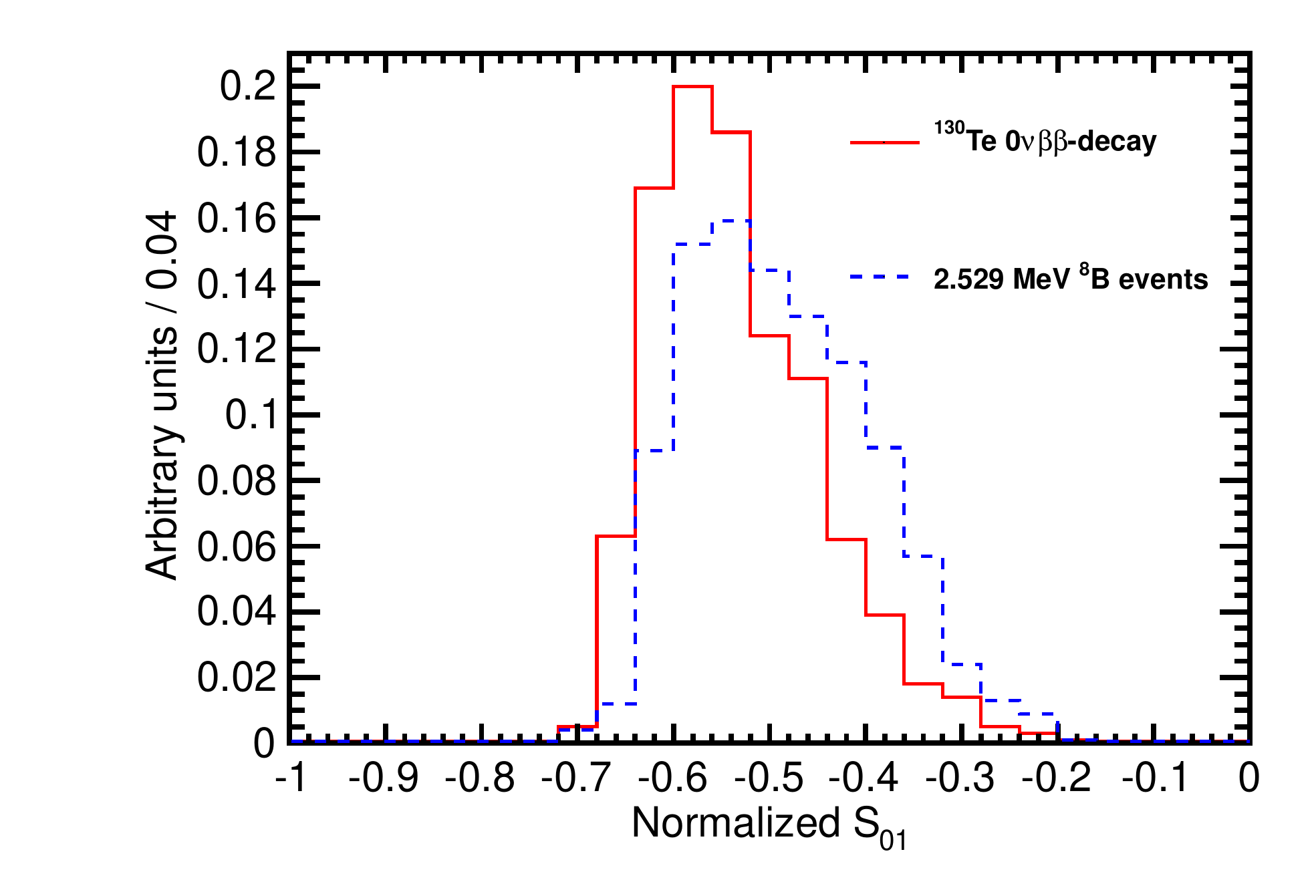}
  \caption{\emph{Left:} Scatter plot of $S_0$ versus $S_1$ for a
    simulation of 1000 signal (\emph{red crosses}) and background
    (\emph{blue triangles}) events. Event vertices are uniformly
    distributed within the fiducial volume, $R<3$~m.  The vertex is
    smeared with 5.2~cm resolution. A differential cut of $\Delta
    t=t^{phot}_{measured} - t^{phot}_{predicted}<$1~ns is applied to
    select the early PE sample.  The default QE and 100\% photo-coverage
    are used in the simulation.  The black dashed line corresponds to a
    linear fit to define the 1-D variable $S_{01}$.
    \emph{Right:} A comparison of the $S_{01}$ distribution between
    signal (\emph{red solid line}) and background (\emph{blue dashed
    line}).  $I_{overlap}$=0.79.}
\label{fig:SL_Te_SmearX3cm_momDT1ns_rndVtx_3p0m}
\end{figure*}

Figure~\ref{fig:SL_Te_SmearX3cm_momDT1ns_rndVtx_3p0m} shows the
performance of the spherical harmonics analysis under these more
realistic assumptions. The overlap between signal and background is
$I_{overlap}$=0.79, which means that the separation is 52\% worse than
in an idealized scenario shown in Fig.~\ref{fig:SL_Te_33p5ns_center}.
The spherical harmonics analysis brings little separation between
signal and background in our default detector model after the
chromatic dispersion and vertex resolution are taken into
account. However, properties of the liquid scintillator can be
adjusted to improve the performance of the spherical harmonics
analysis. In the following we show that a single change in the
scintillation rise time improves the separation.

\subsection{Importance of the liquid scintillator properties}
The strong dependence on the vertex resolution can be addressed by
choosing a liquid scintillator mixture with a more delayed emission of
scintillation light with respect to Cherenkov light. With a larger
delay in scintillation light, a higher fraction of Cherenkov light can
be maintained in the early PE sample even if 
the vertex position is mis-reconstructed. 
In addition, if the fraction of scintillation light is small
compared to Cherenkov light, the distortions in the uniformity of the
scintillation PE due to a shifted reconstructed vertex position does
not significantly affect the spherical harmonics power
spectrum. Furthermore, the effects due to chromatic dispersion can be
addressed by using liquid scintillators with a narrower emission
spectrum~\cite{Aberle2014}, or red-enhanced
photocathodes~\cite{Aberle2014}.

\begin{figure*}[h]
  \centering
  \includegraphics[width=0.49\textwidth]{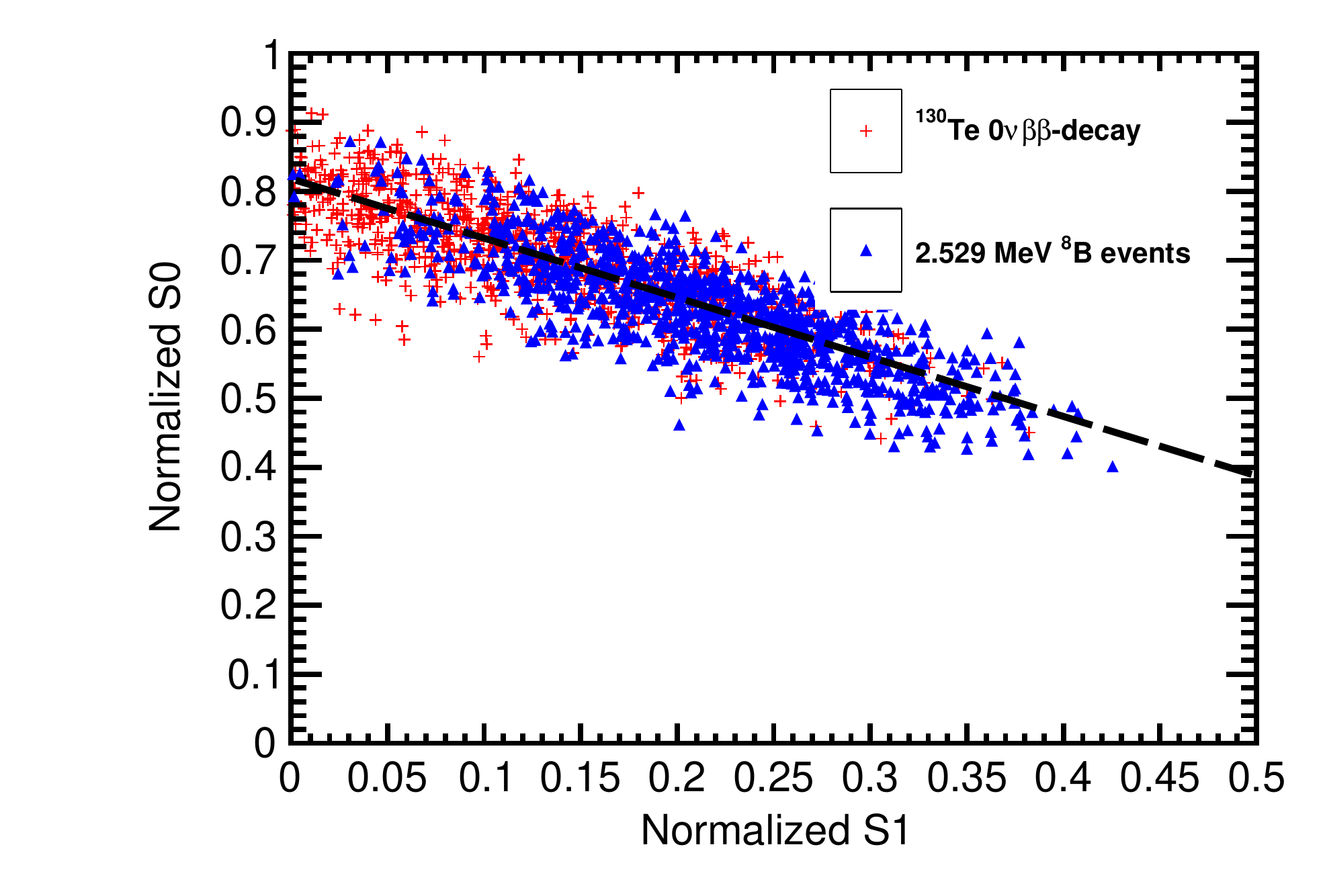} 
  \includegraphics[width=0.49\textwidth]{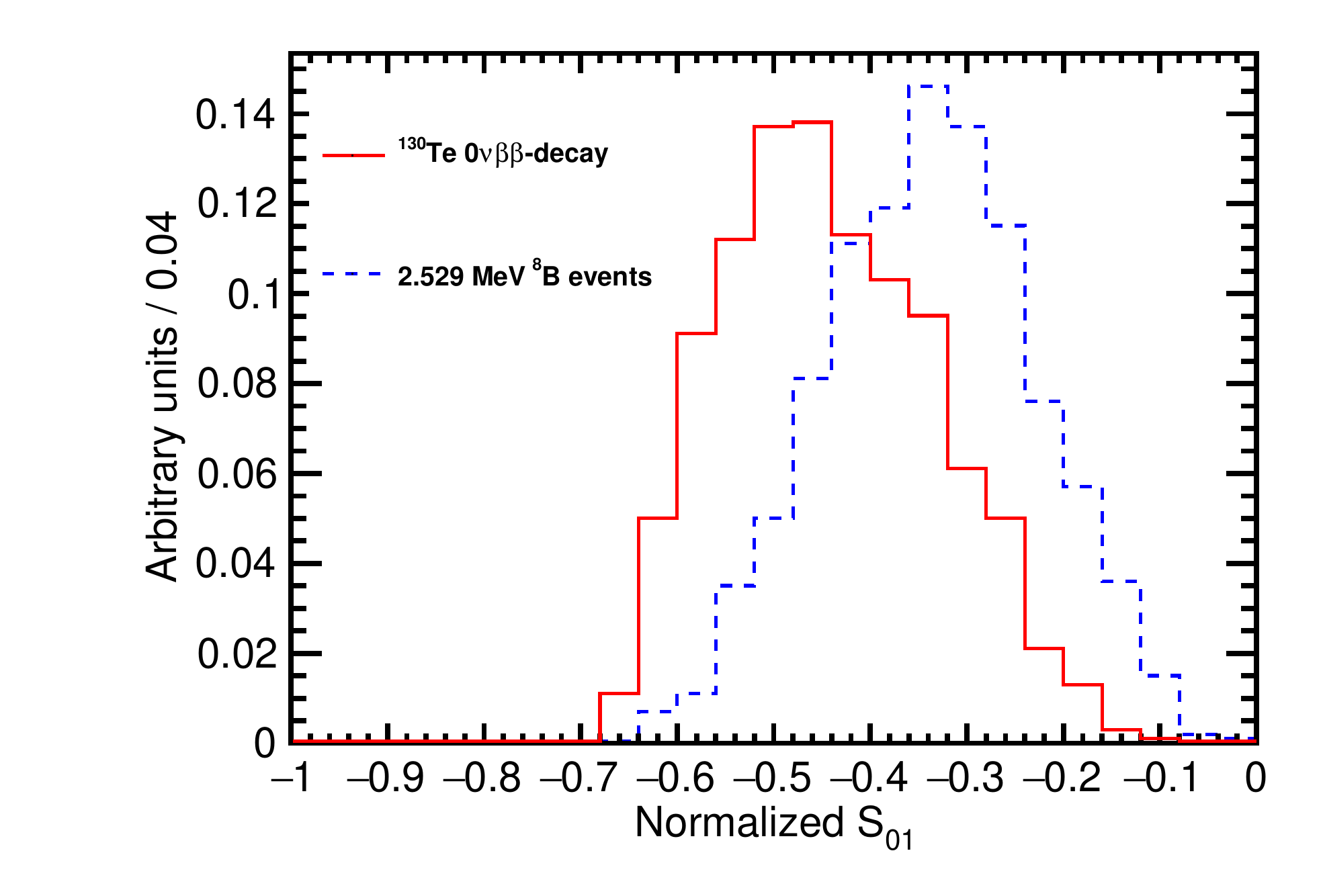}
  \caption{The scintillation rise time constant is increased to
    $\tau_r=$5~ns compared to $\tau_r=$1~ns in the default detector
    model.  \emph{Left:} Scatter plot of $S_0$ versus $S_1$ for a
    simulation of 1000 signal (\emph{red crosses}) and background
    (\emph{blue triangles}) events. Event vertices are uniformly
    distributed within the fiducial volume, $R<3$~m.  Vertex is
    smeared with 5.2~cm resolution. Differential cut of $\Delta
    t=t^{phot}_{measured} - t^{phot}_{predicted}<$1~ns is applied to
    select early PE sample.  The default QE and 100\% photo-coverage
    is used in the simulation.  Black dashed line corresponds to a
    linear fit to define 1-D variable $S_{01}$ (see text for details).
    \emph{Right:} Comparison of the $S_{01}$ distribution between
    signal (\emph{red solid line}) and background (\emph{blue dashed
    line}).  $I_{overlap}$=0.64.}
\label{fig:SL_Te_SmearX3cm_momDT1ns_rndVtx_3p0m_SciRT5p0ns}
\end{figure*}

While the default detector model assumes a scintillation rise time of
$\tau_r=$1~ns, rise times up to $\tau_r=$7~ns can be achieved (see
Ref.~\cite{Minfang_slow_rise_time}). As a test we increased the
scintillation rise time parameter to $\tau_r=$5~ns in the detector
model, with all other parameters kept the same.\footnote{Usually,
longer rise time implies lower light yeild. Here we keep exactly
the same light yeild as in the default detector model, assuming future
possible advances in liquid scintillator technology~\cite{Minfang_private}.} 
Figure
~\ref{fig:SL_Te_SmearX3cm_momDT1ns_rndVtx_3p0m_SciRT5p0ns} shows the
overlap between signal and background is significantly decreased to
$I_{overlap}$=0.64, i.e. the separation is 23\% worse than in the
idealized scenario shown in Fig.~\ref{fig:SL_Te_33p5ns_center} and
23\% better than in the default detector model shown in
Fig.~\ref{fig:SL_Te_SmearX3cm_momDT1ns_rndVtx_3p0m}.

\clearpage

\begin{figure*}[h]
  \centering
  \includegraphics[width=0.49\textwidth]{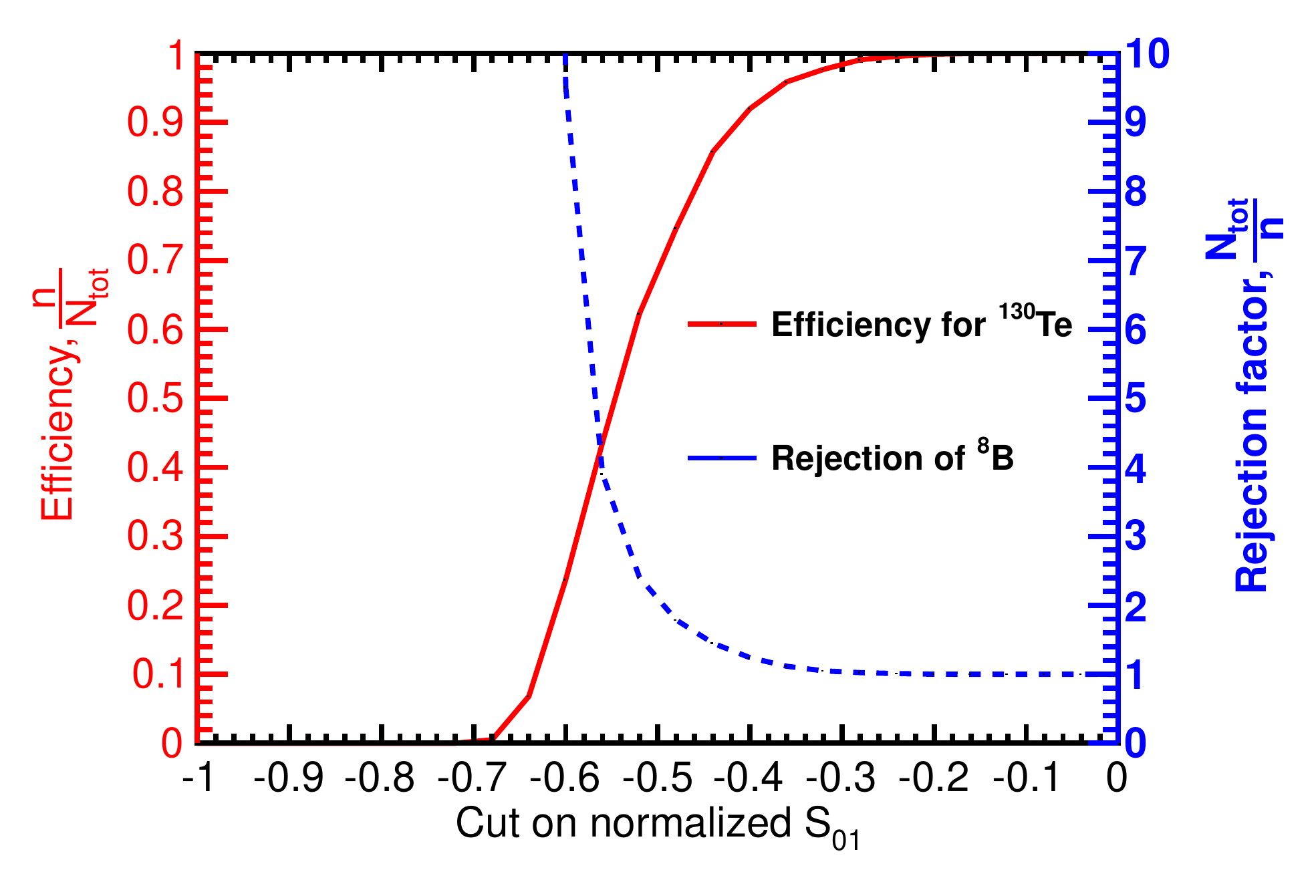}
  \includegraphics[width=0.49\textwidth]{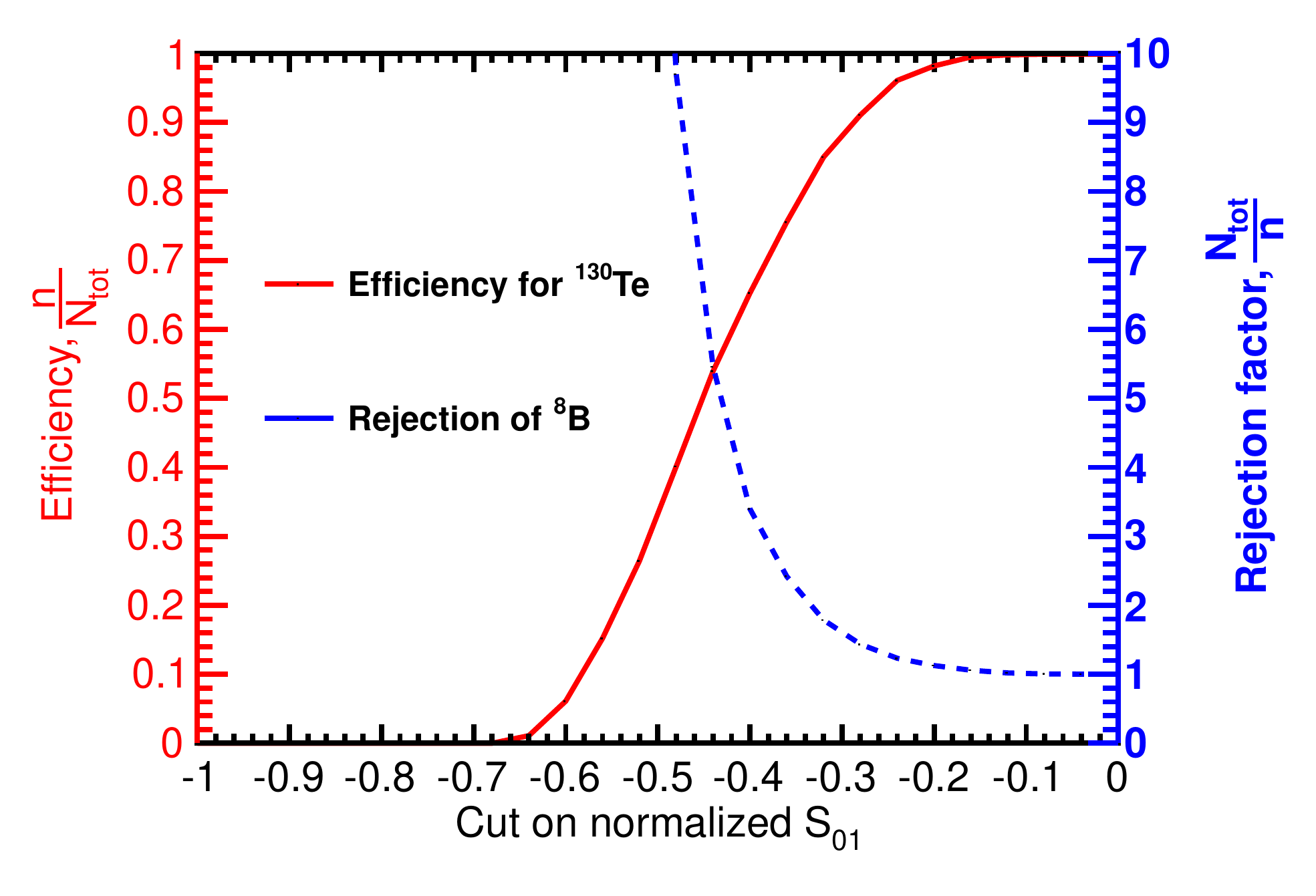}
  \caption{ The efficiency for 0\nbb-decay signal (\emph{red left-hand-scale}) and the
rejection factor for \B~neutrino background (\emph{blue right-hand-scale}) versus
$S_{01}$ for the default model (\emph{left panel}) and a
liquid scintillator with a 5-nsec risetime (\emph{right panel}).}
\label{fig:Efficiency_Rejection}
\end{figure*}

Figure~\ref{fig:Efficiency_Rejection} shows the
efficiency for 0\nbb-decay signal and the rejection factor for \B~ neutrino
background for the default model (left-hand panel) 
and for the slower scintillator with
a 5-ns risetime (right-hand panel) as a function of the $S_{01}$
discriminant. We find a rejection factor of 2 for the default case at
70\% efficiency for signal. The rejection is increased to a factor of
3 for the 5-nsec risetime scintillator.

\section{Conclusions}
\label{sec:conclusions}
We consider the use of large-area photodetectors with good time and
space resolution in kiloton scale liquid scintillator detectors to
suppress background coming from $^{8}$B solar neutrino
interactions. Using a default model detector with parameters derived
from present practice, we show that a sample of detected photons
enriched in Cherenkov light by a cut on time-of-arrival contains
directional information that can be used to separate 0\nbb-decay from
$^{8}$B solar neutrino interactions. The separation is based on a
spherical harmonics analysis of the event topologies of the
two electrons in signal events and the single electron in the
background. The performance of the technique is constrained by
chromatic dispersion, vertex reconstruction, and the time profile of
the emission of scintillation light. The development of a scintillator
with a rise time constant of at least 5~ns would allow a
Cherenkov-scintillation light separation with a background rejection
factor for \B~ solar neutrinos of 3 and an efficiency for 0\nbb-decay signal
of 70\%.

\section*{Acknowledgements}
The activities at the University of Chicago  were supported by the
Department of Energy under DE-SC-0008172, the
National Science Foundation under grant PHY-1066014, and the Driskill
Foundation, and at MIT by  the
National Science Foundation under grant 1554875.

We thank G. Orebi Gann for a discussion on expected backgrounds
at the SNO+ experiment, and J.  Kotilla for discussions on electron
angular correlations in 0\nbb-decay and for providing data with phase
factors for generating 0\nbb- and 2\nbb-decay events.  We are grateful
to C. Aberle for initial development of the Geant-4
detector model used in this paper and for contributions to the
development of the Cherenkov/scintillation light separation technique, and
to M.  Wetstein for help with vertex reconstruction algorithms and
productive discussions on Cherenkov/scintillation light separation. We
thank E. Spieglan for productive discussions on spherical harmonics
analysis and E. Angelico for estimating the effects of photo-detector
position and time resolution on the vertex reconstruction and
verifying the effects of chromatic dispersion.  We thank J. Flusser
for helpful discussions on image processing using moment
invariants. Last but not least we thank M. Yeh for discussions of the
timing properties of liquid scintillators.


\clearpage

\appendix
\renewcommand*{\thesection}{\Alph{section}}
\section{Appendix A}
\label{Appendix_A}

\subsection{Defining the Power Spectrum}

Let the function $f(\theta,\phi)$ represent the distribution of the
photo-electrons (PE) on the detector surface. The function
$f(\theta,\phi)$ can be decomposed into a sum of spherical harmonics:

\begin{eqnarray}
\label{eq1}
f(\theta,\phi) = \sum_{\ell=0}^{\infty} \sum_{m=-\ell}^{\ell} f_{\ell m} Y_{\ell m}(\theta,\phi),
\end{eqnarray}

where $Y_{\ell m}$ are Laplace's spherical harmonics defined in a
real-value basis using Legendre polynomials $P_{\ell}$~\cite{legendre_polynomials}:

\begin{eqnarray}
\label{eq2}
Y_{\ell m} = \left\{
  \begin{array}{@{}ll@{}}
    \sqrt{2}N_{\ell m}P_{\ell}^m(\cos\theta)\cos~m\phi, & \text{if}\ m>0 \\
    N_{\ell m} = \sqrt{\frac{(2\ell+1)}{4\pi} \frac{(\ell-m)!}{(\ell+m)!}}, & \text{if}\ m=0 \\
    \sqrt{2}N_{\ell |m|}P_{\ell}^{|m|}(\cos\theta)\sin~|m|\phi, & \text{if}\ m<0
  \end{array}\right.
\end{eqnarray}

where the coefficients $f_{\ell m}$ are defined as
 
\begin{eqnarray}
\label{eq3}
f_{\ell m} = \int_{0}^{2\pi} d\phi \int_0^{\pi} d\theta \sin\theta f(\theta,\phi) Y_{\ell m}(\theta,\phi).
\end{eqnarray}

Equation~\ref{eq4} defines the power spectrum of $f(\theta,\phi)$ in
the spherical harmonics representation, $s_{\ell}$, where $l$ is a multipole
moment. The power spectrum, $s_{\ell}$, is invariant under rotation. 

\begin{eqnarray}
\label{eq4}
s_{\ell} = \sum_{m=-\ell}^{m=\ell} |f_{\ell m}|^2
\end{eqnarray}

The event topology in a spherical detector determines the distribution
of the PE's on the detector sphere, and, therefore, a set of
$s_{\ell}$'s. These values can serve as a quantitative figure of merit for
different event topologies. The rotation invariance of the $s_{\ell}$'s ensures
that this figure of merit does not depend on the orientation of the
event with respect to the chosen coordinate frame.

The sum of $s_{\ell}$'s over all multipole moments equals to the $L^2$ norm of the
function $f(\theta,\phi)$:

\begin{eqnarray}
\label{eq5}
\sum_{\ell=0}^{\infty} s_{\ell} = \int_{\Omega} |f(\theta,\phi)|^2 d\Omega.
\end{eqnarray}

The normalized power spectrum is thus:

\begin{eqnarray}
\label{eq6}
\mathcal{S}_{\ell} = \frac{s_{\ell}}{\sum_{\ell=0}^{\infty} s_{\ell}} =  \frac{s_{\ell}}{\int_{\Omega} |f(\theta,\phi)|^2 d\Omega},
\end{eqnarray}

and can be used to compare the shapes of various functions
$f(\theta,\phi)$ with different normalizations. As the total number of
PEs detected on the detector sphere fluctuates from event to event we
use the normalized power $\mathcal{S}_{\ell}$.

\subsection{Spherical Harmonics Analysis and Off-center Events}

\begin{figure*}[h]
  \centering
  \includegraphics[width=0.4\textwidth]{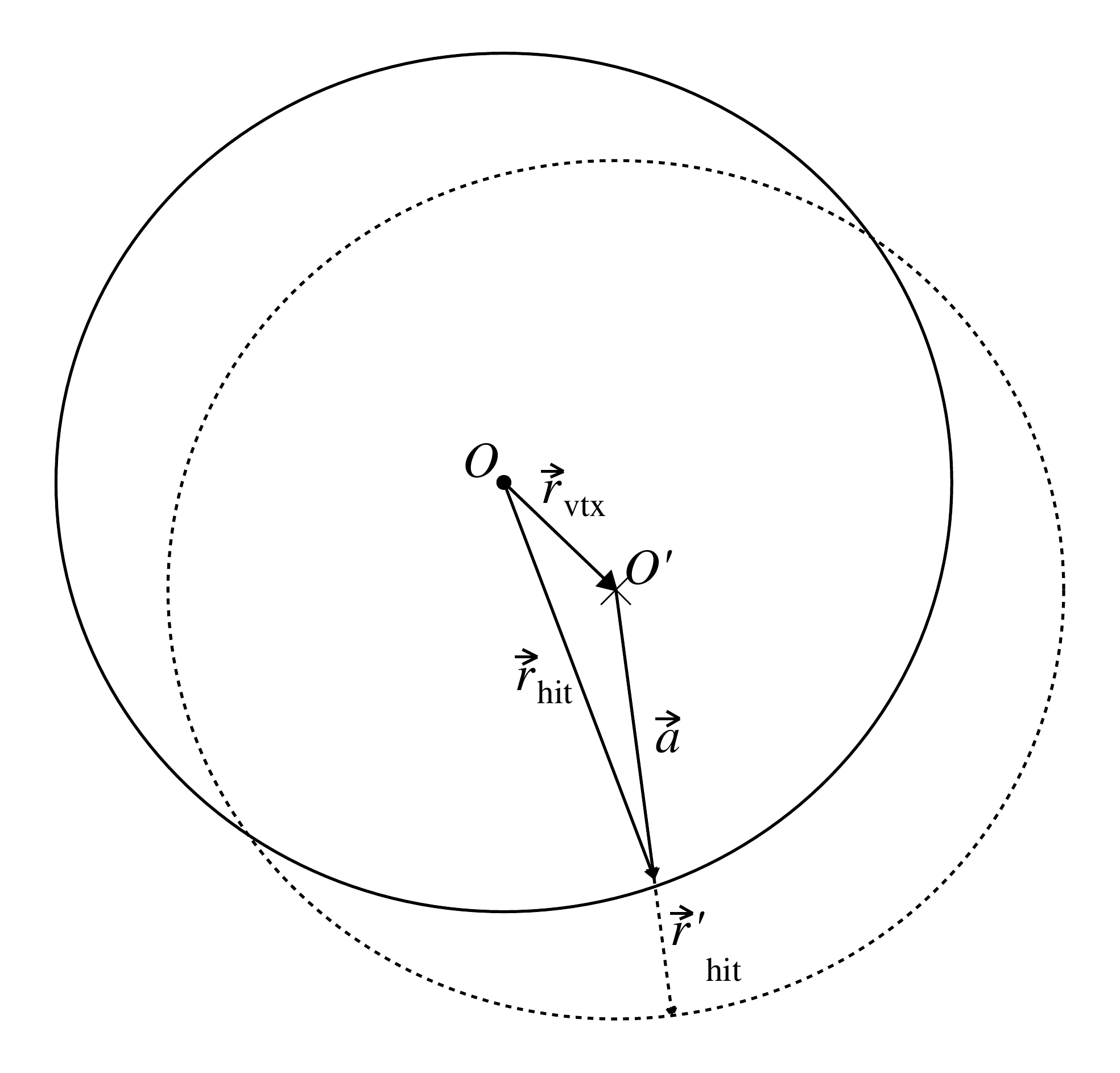}
  \caption{The coordinate transformation which is applied to events that are
    off-center. The solid circle schematically shows the actual detector
    boundaries. The dotted circle shows a new sphere of radius R$=$6.5~m
    with the event vertex position in the center. The radius vector of
    each photon hit is stretched or shortened until the intersection with
    this new sphere using the transformation $\vec{r}^{,}_{hit} =
    \frac{\vec{a}}{|\vec{a}|} \cdot R$, where $\vec{r}^{,}_{hit}$ is a
    new radius vector of the photon hit, $R$ is detector sphere radius,
    and $\vec{a}=\vec{r}_{hit} - \vec{r}_{vtx}$ with $\vec{r}_{hit}$
    and $\vec{r}_{vtx}$ being the radius vectors of the photon hit and
    vertex position in original coordinates, respectively.}
  \label{fig:SphH_transform}
\end{figure*}

In general, events with the same event topology result in the same
the power spectrum $S_{\ell}$ only if events originate in the center 
of the detector. In
order to compare the spherical harmonics for events with verticies away
from the center, a coordinate transformation for each photon hit is
needed. The necessary transformation applied for each PE within an
event is illustrated in Fig.~\ref{fig:SphH_transform}.  The solid
circle in Fig.~\ref{fig:SphH_transform}~has a radius R and shows the
actual detector boundaries. The dotted circle shows a new sphere with
the same radius R, which now has the event vertex in its center. The
radius vector of each PE is stretched or shortened to its intersection
with this new sphere using the transformation, $\vec{r}^{,}_{PE} =
\frac{\vec{a}}{|\vec{a}|} \cdot R$, where $\vec{r}^{,}_{PE}$ is a new
radius vector of a PE and $\vec{a}=\vec{r}_{PE} - \vec{r}_{vtx}$ with
$\vec{r}_{PE}$ and $\vec{r}_{vtx}$ being radius vectors of the PE and
the vertex in the original coordinates, respectively.

\clearpage

\bibliography{bibliography}


\end{document}